\RequirePackage{fix-cm}
\documentclass[smallextended]{svjour3}       
\smartqed  
\usepackage{marvosym}
\usepackage{ifsym}
\usepackage{graphicx}
\usepackage{microtype}
\usepackage{subfig}
\usepackage{graphicx}
\usepackage{graphicx}
\usepackage{booktabs} 

\usepackage{adjustbox}
\usepackage{float}

\usepackage{mathtools}
\usepackage{color}
\usepackage{soul}
\usepackage{amsmath}
\usepackage{amssymb}

\usepackage{hyperref}
\usepackage[numbers,square]{natbib}

\usepackage{authblk}

\begin{document}

\title{\textbf{To BAN or Not to BAN:\\ Bayesian Attention Networks for \\ Reliable Hate Speech Detection}}

\titlerunning{Bayesian Attention Networks for Reliable Hate Speech Detection}        

\author{Kristian Miok\thanks{Corresponding author} \and Bla\v z \v Skrlj \and Daniela Zaharie \and Marko Robnik-\v{S}ikonja}


\institute{Kristian Miok (\Letter) 
                \at
              West University of Timisoara, Computer Science Department \\
              Bulevardul Vasile Pârvan 4, 300223 Timisoara, Romania\\
           \email{kristian.miok@e-uvt.ro}
           \and
           Bla\v z \v Skrlj \at
              Jo\v{z}ef Stefan International Postgraduate School, and Jo\v{z}ef Stefan Institute,\\ Jamova 39, 1000 Ljubljana, Slovenia \\  
                \email{blaz.skrlj@ijs.si}
              \and
          Daniela Zaharie \at
              West University of Timisoara, Computer Science Department \\
              Bulevardul Vasile Pârvan 4, 300223 Timisoara, Romania\\
              \email{daniela.zaharie@e-uvt.ro}  
          \and
          Marko Robnik-\v{S}ikonja \at
          University of Ljubljana, Faculty of Computer and Information Science,\\ Ve\v{c}na pot 113, 1000 Ljubljana, Slovenia \\
        \email{marko.robnik@fri.uni-lj.si} 
}


\maketitle

\begin{abstract}

\textbf{Background}
Hate speech is an important problem in the management of user-generated content. To remove offensive content or ban misbehaving users, content moderators need reliable hate speech detectors. Recently, deep neural networks based on the transformer architecture, such as the (multilingual) BERT model, achieve superior performance in many natural language classification tasks, including hate speech detection. So far, these methods have not been able to quantify their output in terms of reliability.
\\
\\
\textbf{Methods} We propose a Bayesian method using Monte Carlo dropout within the attention layers of the transformer models to provide well-calibrated reliability estimates. We evaluate and visualize the results of the proposed approach on hate speech detection problems in several languages. Additionally, we test if affective dimensions can enhance the information extracted by the BERT model in hate speech classification.
\\
\\
\textbf{Results} Our experiments show that Monte Carlo dropout provides a viable mechanism for reliability estimation in transformer networks. Used within the BERT model, it offers state-of-the-art classification performance and can detect less trusted predictions. 
Also, it was observed that affective dimensions extracted using sentic computing methods can provide insights toward interpretation of emotions involved in hate speech.
\\
\\
\textbf{Conclusions} Our approach not only improves the classification performance of the state-of-the-art multilingual BERT model but the computed reliability scores also significantly reduce the workload in an inspection of offending cases and reannotation campaigns. The provided visualization helps to understand the borderline outcomes.

\end{abstract}

\section{Introduction}
\label{submission}
With the rise of social network popularity, hate speech phenomena have significantly increased \citep{davidson2017automated}. Hate speech not only harms both minority groups and the whole society, but it can lead to actual crimes \citep{bleich2011rise}. Thus, (automated) hate speech detection mechanisms are urgently needed. However, falsely accusing people of hate speech is also a problem. Many content providers rely on human moderators to reliably decide if a given text is offensive or not, but this is a mundane and stressful job which can even cause post-traumatic stress disorders\footnote{\url{https://www.bbc.com/news/technology-51245616}}. There have been many attempts to automate the detection of hate speech in social media using machine learning, but existing models lack the quantification of reliability for their decisions. 

In the last few years, recurrent neural networks (RNNs) were the most popular text classification choice. Long Short Term Memory (LSTM) networks, the most successful RNN architecture, were already successfully adapted for the assessment of predictive reliability in hate speech classification  \citep{miok2019prediction}. Recently, neural network architecture with attention layers, called `transformer architecture' \cite{vaswani2017attention}, showed even better performance on almost all language processing tasks. Using transformer networks for masked language modeling produced breakthrough pretrained models, such as BERT (Bidirectional Encoder Representations from Transformers) \citep{devlin2019bert}. The attention mechanism, which is a crucial part of transformer networks, became an essential part of natural language understanding with a significant impact on language applications. We aim to investigate the behavior of the attention mechanism concerning the reliability of predictions. We focus on the hate speech recognition task. 

In hate speech detection, reliable predictions are needed to remove harmful content and possibly ban malicious users without harming the freedom of speech \cite{miok2019prediction}. Standard neural networks are inadequate for the assessment of predictive uncertainty, and the best solution is to use the Bayesian inference framework. However, classical Bayesian inference techniques do not scale well in neural networks with high dimensional parameter space \cite{izmailov2019subspace}. Various methods were proposed in order to overcome this problem \cite{myshkov2016posterior}. One of the most efficient methods is called Monte Carlo Dropout (MCD) \cite{gal2016dropout}. Its idea is to use dropout in neural networks as a regularization technique \cite{srivastava2014dropout} and interpret it as a Bayesian optimization approach that takes samples from the approximate posterior distribution.

Several authors have shown that emotional information \citep{cambria2016affective} extracted from a text can improve the performance of lexical approaches and standard machine learning algorithms \citep{martins2018hate,alorainy2018suspended,rodriguez2019automatic,bauwelinck2019measuring}. The role and utility of emotional information  in deep learning have not yet been established; besides, we still have only limited understanding of the emotions in the text. A series of computational models that bridge the gap between the human emotional perspective evolved in a domain known as 'Sentic Computing' \cite{cambria2012sentic}. The computational initiative, named 'SenticNet', combines knowledge from psycholinguists, neuroscientists, and computer scientists to better understand emotions in text. We used information on affective dimensions provided by SenticNet, together with the outputs of the state-of-the-art contextual language model BERT \citep{devlin2019bert}. This was enhanced with a reliability estimation mechanism based on MCD as input for a hate speech classifier. Concerning emotions, we follow two goals in this work: i) to test the predictive performance of emotion-enhanced BERT models in hate speech detection, and ii) to better understand the role of emotions in hate speech.


Our main contributions are:
\begin{enumerate}
\item We present a novel methodology for the assessment of prediction uncertainty in attention networks and in BERT models.
\item Empirical analysis of the proposed Bayesian Attention Networks (BANs) and MCD enhanced BERT models show an improved calibration and prediction performance on hate speech detection tasks in several languages.
\item We combine contextual and reliability information obtained from MCD BERT with sentiment-related knowledge provided by SenticNet.

\item We demonstrate novel visualization of prediction uncertainty for individual instances, as well as for groups of instances. 
\end{enumerate}

The paper consists of six more sections. In Section 2, we present related works on prediction uncertainty, hate speech detection and its relationship with sentiment analysis. In Section 3, we propose the methodology for uncertainty assessment in transformer networks using attention layers and MCD, while in Section 4, we analyze the calibration of predictions. Section 5 presents the datasets and the evaluation scenario. The obtained results are presented in Section 6, followed by conclusions and ideas for further work in Section 7.

\section{Related Work}
We present the related work categorized into four areas. In Section \ref{sec:relatedHate}, we introduce work done on hate speech
detection, followed by the related research on transformer architecture for text classification in Section \ref{sec:relatedAttention}. In Section \ref{sec:relatedUncertainty}, we describe existing approaches for the assessment of uncertainty in text classification. Finally, in Section \ref{sec:relatedSentic}, we relate hate speech detection with the particularities of sentic computing.

\subsection{Hate Speech Detection}
\label{sec:relatedHate}

Analyzing sentiments and extracting emotions from texts are very useful natural language processing (NLP) applications. With the rise of social media popularity, the hate speech detection became highly needed. Hate speech is defined as written or oral communication that abuses or threatens a specific group or target \cite{warner2012detecting}. 

Detecting abusive language for less-resourced languages is complex, and has inspired research in multilingual and cross-lingual methods \cite{stappen2020cross}. These methods are especially useful when the involved languages are morphologically or geographically close \cite{pamungkas2019cross}. In our work, we investigate hate speech detection methods for English, Croatian, and Slovene languages. The English language is well-resourced and researched \cite{malmasi2017detecting,davidson2017automated,waseem2016hateful}. Recently, hate speech detection studies appeared for Croatian \cite{kocijan2019detecting,marinsek2019cross,ljubevsic2018datasets} and Slovene \cite{fivser2017legal,ljubevsic2019frenk,vezjak2018radical}.  

The hate speech detection is mostly treated as a binary text classification problem.  In the past, the most frequently used classifier was the Support Vector Machines (SVM) method \cite{schmidt2017survey}. However, deep neural networks are now a dominant technique, first through RNNs \cite{mehdad2016characters}, and recently using the pre-trained transformer networks  \cite{mozafari2019bert,wiedemann2020uhh}. In this work, we analyzed the state-of-the-art pre-trained  transformer networks, called (multilingual) BERT model.

\subsection{Attention Networks for Text Classification}
\label{sec:relatedAttention}
Attention mechanism is a key component of transformer architecture, proposed by \citet{vaswani2017attention}. Due to its power and suitability for parallelization, this architecture soon replaced LSTM networks for many NLP tasks. Recently, large pre-trained transformer models have been investigated in the context of text classification tasks. For example, \citet{kant2018practical} trained both multiplicative LSTM (mLSTM) and transformer language models on a large 40GB text dataset \cite{mcauley2015image} and transferred those models to binary and multi-class text classification problems. They concluded that the transformer model outperforms the mLSTM model, especially when fine-tuned for multidimensional emotions classification. 


The BERT model \cite{devlin2019bert} uses the transformer architecture and large text corpora to learn masked language model and sequence of sentences tasks.  BERT and its follow-ups are able learn and extract many language characteristics (both syntactic and semantic) and excel for many text classification tasks. Despite the short time since its conception, BERT has already attracted enormous attention from the NLP community. Hundreds of research groups extensively research it; see a recent overview by \citet{rogers2020primer}. Practical guidelines on how to fine-tune the BERT model for text classification were compiled by \citet{sun2019fine}.

A multilingual hierarchical attention mechanism for document classification was investigated by several authors \cite{pappas2017multilingual,du2017stance,yang2016hierarchical}. However, different attention layers of large pre-trained models were not tested separately or in the context of prediction reliability. Also, to the best of our knowledge, the predictive reliability of BERT outputs has not been investigated, yet.

\subsection{Prediction Uncertainty for Text Classification}
\label{sec:relatedUncertainty}

While recent works on classification reliability mostly investigate deep neural networks, many other probabilistic classifiers were analyzed in the past \citep{niculescu2005predicting}. For example, \citet{Platt99probabilisticoutputs} explores the probabilistic properties of SVM predictions. 

Prediction uncertainty is an important issue for black-box models like neural networks, as they do not provide  interpretability or reliability information about their predictions. Most reliability scores for deep neural networks are based on a Bayesian framework. The most popular exception is the work of \citet{lakshminarayanan2017simple}, who proposed using deep ensembles to estimate the prediction uncertainty. 

An efficient approach to reliability assessment in neural networks is to mimic the Bayesian inference using MCD \cite{gal2016dropout}. The dropout technique was first introduced to RNNs in 2013 \cite{wang2013fast}, but further research revealed a negative impact of dropout in RNNs \cite{bluche2015apply}. Later, dropout was successfully applied to language modeling by \citet{zaremba2014recurrent}, who applied it only to fully connected layers. \citet{gal2016theoretically} implemented the variational inference based dropout, which can regularize also recurrent layers. Additionally, they provide a solution for dropout within word embeddings. The method mimics Bayesian inference by combining probabilistic parameter interpretation and deep RNNs. The authors introduce the idea of augmenting probabilistic RNN models with the prediction uncertainty estimation. Several other works investigate how to estimate prediction uncertainty using RNNs \cite{zhu2017deep}, e.g., Bayes by Backpropagation (BBB) \citep{fortunato2017bayesian}. 

Recently, a fast and scalable method called `SWAG' was proposed by \citet{maddox2019simple}. The main idea of this method is to randomize the learning rate and interpret it as a sampling from the Gaussian distribution. SWAG fits the Gaussian distribution by capturing the Stochastic Weight Averaging (SWA) mean and co-variance matrix, representing the first two moments of stochastic gradient descent iterations. Different to SWAG, we use the Gaussian distribution as a posterior over neural network weights, and then perform a Bayesian model averaging for uncertainty estimation and calibration.

MCD was recently used within several models and different architectures to obtain the prediction uncertainty and improve the classification results \cite{inproceedings,miok2019generating,miok2020multiple}. Transformer networks were not yet analyzed. 

\subsection{Sentic Computing}
\label{sec:relatedSentic}

Sentiments and emotions play an essential role in hate and offensive speech, and have been used successfully in their automatic detection. \citet{martins2018hate} have used eight basic emotions from Plutchik's model \cite{Plutchik1980emotion}, the positive and negative sentiment polarities, indicator of a presence of a word in the Hatebase lexicon\footnote{\url{http://www.hatebase.org}}, and the intensity of anger emotion. Their combination of the lexicon-based and machine learning approach successfully predicted hate speech and showed a high utility of emotional features. Alorainy et al., 2018 \cite{alorainy2018suspended} used the emotional analysis on Twitter suspended accounts and discovered that they contain more disgust, negative sentiment, fear, and sadness than active accounts. Using this information for hate speech detection, their machine learning models showed improved performance. \citet{rodriguez2019automatic} also used the eight basic emotions in their emotional analysis and showed that emotions could improve Facebook posts' clustering. Finally, \citet{bauwelinck2019measuring} used several different groups of features (linguistic, sentiment, and Twitter-specific features such as hashtags and profanity lexicon) to predict hate speech. Interestingly, their results show that Twitter-specific features are the most successful, and the additional sentiment features do not improve predictive performance. 
All the methods mentioned above use either classical machine learning approaches such as SVM, Naive Bayes, logistic regression, and random forest, or RNNs, such as LSTMs.

To advance approaches based on lexical keywords and frequency statistics, \citet{cambria2012sentic} proposed a framework for emotional computing called `SenticNet' that captures semantics and latent emotional information by relying on the implicit meaning associated with commonsense concepts. The original emotion categorization model called Hourglass of Emotions \cite{cambria2012org} was supported by the SenticNet 4 framework \cite{cambria2012sentic}, while its newer revised version \cite{susanto2020hourglass} is used in SenticNet 6 framework \citep{cambria2020senticnet}. These models are biologically-inspired and psychologically-motivated.  Each of the two models is based on four independent but concomitant affective dimensions, which can be combined to build more complex emotions. Based on this, SenticNet framework can describe and explain emotional experiences by disassembling text to the ground sentiments. 

The SenticNet framework has been successfully used in sentiment classification problems. 
 Sentic LSTM \citep{ma2018sentic} integrates the explicit emotional information  with the LSTM networks  by adding a recurrent additive network that simulates sentic patterns. 
A recent SenticNet 6 framework  \citep{cambria2020senticnet} combines top-down and bottom-up knowledge representation. From top-up direction it encodes meaning using symbolic models (logic and semantic
networks); in bottom-up direction it learn syntactic patterns from data, using subsymbolic
methods (biLSTM and BERT). Authors report state-of-the-art results for sentiment analyses. 

In our work, we use the transformer architecture that can extract highly relevant information from texts. Concerning emotions, the question we investigate is whether adding emotional information to the distribution of predictions can improve the performance of hate speech detection. This question is particularly relevant for the current state-of the-art BERT model \citep{devlin2019bert}, which is known to capture a plethora of language information, such as part-of-speech tags, dependency structure, and sentiment.


\section{Bayesian Attention Networks}
\label{sec:BAN}
The BERT model \citep{devlin2019bert} is the transformer network that has achieved state-of-the-art results in many NLP tasks, including text classification \citep{xu2020improving,gururangan2019variational,chang2019x}. In this work, we introduce Monte Carlo Dropout to transformer networks and BERT to construct their Bayesian variants. Analysis of different amounts of dropout, different variants of BERT modifications, and their hyper-parameters would require pretraining and fine-tuning several different BERT models, which would require substantial computational resources. For example, pretraining a single BERT model on four Tensor Processing Units (TPUs) requires more than a month of computational time. Thus, in this work, we  explore two reliability  extensions, i) the reliability on the encoder part of the BERT architecture trained from scratch (without pretraining) on the task of interest (in this work, we refer to these models as the attention networks), and ii) reliability of pre-trained BERT models, using only fine-tuning.  We believe this is a reasonable setting which sheds light on an important reliability aspect of transformer networks. 


In Section \ref{sec:AN}, we first formally define the attention network architecture, and in Section \ref{sec:MCDAN}, we make it Bayesian by introducing MCD. Finally, in Section \ref{sec:MCDBERT}, we describe how the MCD principle can be employed in already pre-trained BERT models.

\subsection{Attention Networks}
\label{sec:AN}
\begin{figure}[htb]
  \centering
     \includegraphics[width=0.4 \linewidth]{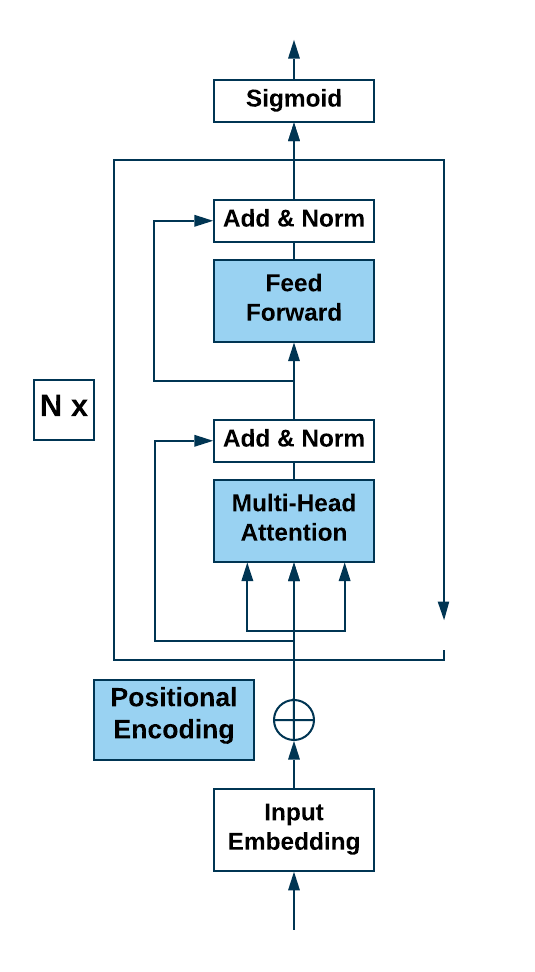}
        \caption{A scheme of Attention Networks. The dropout is introduced in the blue colored layers.
        }
        \label{fig:int}
\end{figure}

The basic architecture of the attention network follows the architecture of transformer networks \cite{vaswani2017attention} and is shown in Figure~\ref{fig:int}. 
The proposed architecture is similar to the encoder part of the transformer architecture. The difference is in the output part, where a single output head was added to perform binary classification using the sigmoid activation function. The main difference to BERT, which also uses just the encoder part of transformer network, is that we do not use any pretraining.  The second difference is that attention network uses the classification head and BERT has the language model head. In both cases the output is composed of feed-forward layers followed by the non-linearity but with different dimensions in each case.
By not relying on the pretraining, we are much more flexible concerning the number of layers and number of neurons in each layer. For our tasks, we use orders of magnitude fewer parameters, e.g., we used a maximum of 3 million parameters (at the expense of loosing information from pretraining). The architecture can contain many attention heads, where a single attention head is computed as:
\begin{equation*}
    o_h = \textrm{softmax}(\frac{\boldsymbol{Q} \cdot \boldsymbol{K}^T}{\sqrt{d_k}})\cdot \boldsymbol{V},
\end{equation*}
The attention matrices are commonly known as the query $\boldsymbol{Q}$, the key $\boldsymbol{K}$, and the value matrix $\boldsymbol{V}$. The normalizing factor, $d_k$, denotes the dimensionality of keys. The attention function can be described as mapping the query and the set of key-value pairs to the output, where the query, keys, values, and output are all vectors. The output is computed as a weighted sum of the values. The weight assigned to each value is computed by a compatibility function of the query with the corresponding key. 

Intuitively, the multiplication of query and key vectors with subsequent values can be understood as the extraction of \emph{relations}. The softmax activation enables each pair of considered input tokens to be represented with a single real value. It effectively introduces \emph{sparseness} into the weight space -- only certain token pairs emerge with high weights and are relevant for the remaining part of the considered neural network architecture.
In practice, multiple such heads can be concatenated and fed into the succeeding feed-forward layer.
The application of softmax has been shown to emphasize only particular parts of the parameter space, thereby making the neural network more focused.

 The positional encoding, as discussed in \citet{vaswani2017attention}, represents a matrix that encodes individual positions in a matrix of the same dimensionality as the one holding the information on sequences (input embedding). The positional encoding was introduced to account for \emph{word order}. Here, relative distances between different tokens are taken into account by incorporating the position-related signal into a given token representation.
 
While there are, in principle, many different ways of how attention networks can be extended with the Bayesian approach, we propose to use the well-established MCD.

\subsection{Monte Carlo Dropout for Attention Networks}
\label{sec:MCDAN}
In our proposal, called Bayesian Attention Networks (BAN), we use MCD within attention networks but contrary to the original dropout setting, the dropout layers are active also during the prediction phase. In this way, the predictions are not deterministic but are sampled from the \emph{learned} distribution, thereby forming an ensemble of predictions. The obtained distribution can be, for example, inspected for higher moment properties and can offer additional information on the uncertainty of a given prediction. 
During the prediction phase, the dropout layers are activated again and the output of a proportion of randomly selected neurons in those layers is set to zero.
A forward pass on such partially activated architecture is repeated for a fixed number of samples, every time dropping  different randomly selected neurons. The results of different passes can be combined to obtain the final prediction, or further inspected as a probability distribution.

\subsection{Monte Carlo Dropout for BERT}
\label{sec:MCDBERT}
MCD was used in the BERT model in the same way as in BAN. MCD can provide multiple predictions of a neural network during the test time, as long as the dropout was used during the training phase \cite{gal2016uncertainty}. Training of neural networks with the dropout distributes the captured information across the network. During the prediction, such a trained neural network is robust. Using the dropout principle, a new prediction is possible in each forward pass. A sufficiently large set of such predictions can be used to estimate the prediction reliability. The BERT model is trained with 10\% of dropout in all of the layers by default, and thus allows for multiple predictions using the described principle. We call this model `MCD BERT'. A  limitation of this approach is that a single dropout rate of 10\% is used during training, while other dropout probabilities might be more suitable for reliability estimation. We leave this analysis for further work.




\section{Calibration of Probabilistic Classifiers}
\label{sec:calibration}
The quality of reliability scores returned by probabilistic classifiers (such as BAN and MCD BERT) is assessed with calibration measures. 
A classifier is calibrated if its output scores are close to actual probabilities in a sense that a class predicted with the score $p$ is correct with the actual probability $p$, i.e. in $p \cdot 100$ percent of cases. Without special calibration approaches, most neural networks are overconfident and overestimate their probabilities. The calibration of a model can be visualized using a calibration plot where the model's prediction accuracy (true probabilities) is plotted against the predicted probabilities (i.e. outputs scores). The perfect calibration manifests itself as a diagonal in the calibration plot (see an example of a calibration plot in Fig. \ref{fig:CalEnglishFinal}).

Since classifiers are typically not perfectly calibrated, we investigated different methods to improve the calibration of used neural networks. We compared several existing calibration methods with a novel approach that combines existing techniques with a method for threshold adaptation. In Section \ref{sec:calibrationMethods}, we describe the existing calibration methods, followed by the proposed threshold adaptation in Section \ref{sec:AT}. 

\subsection{Existing Calibration Methods}
\label{sec:calibrationMethods}
We first formally describe how to obtain calibrated predictions from the reliability scores.
Let $(X, Y)$ be the input space, where $X$ represents the set of predictive variables, and $Y$ is the binary class variable (either 0 or 1). Let $f$ be the predictor (e.g., neural network) with $f(X)= (\hat{Y},\hat{P})$, where $\hat{Y}$ is the binary class prediction, and $\hat{P}$ is its associated confidence score or probability score of correct prediction. 
The calibration of the model $f$ is expressed as:
\begin{equation}
P(\hat{Y}=Y|\hat{P}=\hat{p})=p, 
\label{eq:calibration}
\end{equation}
where $\hat{p}$ is the prediction score from $[0, 1]$ interval, obtained from the predictor $f$. We interpret this score as the probability of a specific outcome, assigned by the model $f$. Probability $p$ is the model's confidence or true probability that model $f$ predicts correctly.
If a model predicts a certain outcome with a high probability, it is desirable that the confidence of this prediction being correct is also high. In the ideal case of perfect calibration $\hat{p} = p$. 

Based on Equation (\ref{eq:calibration}), there are two ways to reduce the calibration error: either to obtain calibrated predictions $\hat{p}$ or to manipulate the prediction threshold in such a way that the predicted outcome $\hat{Y}$ is better calibrated.
To assess the quality of the produced reliability scores, we compare them to results of two calibration methods, Platt's method and Isotonic regression.  

Platt's method \citep{Platt99probabilisticoutputs} learns two scalar parameters $a, b \in \mathbb{R}$ in such way that the prediction $\hat{q} = \sigma(a \hat{p} + b)$ presents a calibrated probability of predicted score $\hat{p}$, and  $\sigma$ is the sigmoid function. To find good values of $a$ and $b$, typically a separate calibration dataset is used. The isotonic regression is a non-parametric form of regression in which we assume that the function is chosen from the class of all isotonic (i.e., non-decreasing) functions \cite{zadrozny2002transforming}. Given the predictions from our classifier  $\hat{p}$, and the true target $y$, the calibrated prediction returned by the isotonic regression is:
$$ \hat{q} = m(\hat{p}) + \epsilon $$ 
where $m$ is a non-decreasing function.

\subsection{Adaptive Threshold}
\label{sec:AT}

We explored the adaptive threshold (AT), which we apply to classification with BANs. During learning, after each weight update phase, we assess the performance of BAN. For each instance in the validation set, we do multiple forward passes with unfrozen dropout layers and store the average of the returned scores as the probability estimate. Once the probability estimates for the validation set are collected, we test several decision thresholds and determine the predictions of each instances. The best-performing threshold w.r.t. a given performance metric (in our case the classification accuracy), is stored together with its performance and weights of the neural network. The obtained performance estimate can also be used for early stopping in the learning phase. When we apply the model to new instances, we use the best threshold from the training phase (instead of the default value of 0.5). The purpose of AT is to automatically find the threshold with the best performance. 
To summarize, we employ the following procedure:
\begin{enumerate}
    \item During the training and after each weight update, we generate the probability distribution with MCD. The mean of the distribution is considered the probability score of a given instance being assigned to the positive class.
    
    \item Using the validation set, we test a range of possible thresholds that determine the instances' labels. We tested the threshold range between 0.1 and 0.9 in increments of 0.001.
    
    \item If the accuracy obtained by the default threshold (0.5) was improved by any other threshold, we stored both the current parameter set and the threshold value used to obtain the improved performance on the validation set.
    
    \item The weights of the best performing model and the matching threshold are returned as the final prediction model.
\end{enumerate}

\section{Evaluation Settings}
In this section, we present the evaluation settings, and in Section \ref{sec:results}, we report the results. Starting with Section \ref{sec:datasets}, we describe the used hate speech datasets, followed by he affective dimensions of the Hourglass of Emotions method in Section \ref{sec:senticdata}. The implementation details of the used prediction models are presented in Section \ref{sec:predictionModels}. In Section \ref{sec:evaluationMeasures}, we present the evaluation measures for the predictive performance, and in Section \ref{sec:calibrationEvaluation}, the measures used in the evaluation of calibration.

\subsection{Hate Speech Datasets}
\label{sec:datasets}
To test the proposed methodology in the multilingual context, we used hate speech datasets in three languages, English, Croatian, and Slovene. The summary of datasets is available in Table \ref{table:inputs}.

\begin{enumerate}
    \item The \textbf{English} dataset\footnote{\url{https://github.com/t-davidson/hate-speech-and-offensive-language}} is extracted from hate speech and offensive language detection study of \citet{davidson2017automated}. The subset of data we used consists of 5,000 tweets. We took 1,430 tweets labeled as the hate speech and randomly sampled 3,670 tweets from the remaining 23,353 tweets. 
    
    \item The \textbf{Croatian} dataset was provided by the Styria media company within the EMBEDDIA project\footnote{\url{http://embeddia.eu}}. The texts consists of user comments on the news portal Ve\v{c}ernji list\footnote{\url{https://www.vecernji.hr}}. The original dataset consists of 9,646,634 comments from which we selected 8,422 comments. 50\% of instances were labeled as the hate speech by human moderators, and the other half was chosen randomly from non-problematic comments.
    
    \item  The \textbf{Slovene} dataset was produced in the Slovenian national project FRENK\footnote{\url{http://nl.ijs.si/frenk/} (Research on  Inappropriate Electronic Communication)}. The text dataset used in the experiment is a combination of two different studies of Facebook comments \cite{ljubevsic2019frenk}. The first group of comments was collected on LGBT homophobia topics, while the second on anti-migrants posts. In our final dataset, we used all of the 2,182 hate speech comments, and the same number of non-hate speech comments were randomly sampled. 
\end{enumerate}

\begin{table}[htb]
\caption{Characteristics of the datasets used in the experiments.  }
\label{tab:models}
\renewcommand{\arraystretch}{1}
\setlength{\tabcolsep}{5pt}
\label{table:inputs}
\centering
\begin{tabular}{lccccc}
    \textbf{Dataset} & \textbf{type} & \textbf{Size}  &  \textbf{Hate} & \textbf{Non-hate} & \textbf{LSTM embeddings}  \\
    \hline
    \textbf{English} & tweets & 5000 & 1430 & 3670 & sentence \\
    \textbf{Croatian} & news comments & 8422 & 4211 & 4211 & fastText \\
    \textbf{Slovene} & Facebook comments & 4364 & 2182 & 2182 & fastText \\
     \hline
\end{tabular}
\end{table}

\subsection{The Hourglass of Emotions Affective Dimensions} 
\label{sec:senticdata}

To test if emotional information extracted from text can complement the information extracted by BERT models, we used the English tweets dataset and affective dimensions obtained with two versions of the Hourglass of Emotions model; the affective dimensions of the original model can be extracted using the \textit{SenticNet 4} framework \cite{cambria2012org}, and the affective dimensions of  its revision \cite{susanto2020hourglass} are available in the \textit{SenticNet 6} framework.

\subsubsection{SenticNet 4}
We used the SenticPhrase interface to obtain the original Hourglass of Emotions affective dimensions from the SenticNet 4 framework \cite{cambria2016senticnet}. For each sentence, we extracted four affective dimensions (pleasantness, attention, sensitivity, and aptitude). Within SenticNet 4, verb and noun concepts are linked to primitives, and in this way, most concept inflections can be captured by the knowledge base verb concepts. The implementation is freely accessible via Python API (Application Programming Interface) in the Python sentic package\footnote{\url{https://pypi.org/project/sentic/}}. 

To gain a better understanding of the four affective dimensions, \citet{cambria2010not} presented the following example:
\begin{enumerate}
    \item The user is happy with the service provided (pleasantness).
    \item The user is interested in the information supplied (attention).
    \item The user is comfortable with the interface (sensitivity).
    \item The user is disposed to use the application (aptitude).
\end{enumerate}
The hate speech texts usually express unhappiness with the current situation and unwillingness to hear or consider different opinions. Hence, the nature of the hate speech is opposite to the nature of pleasantness and aptitude, while it can be correlated with the attention. 

\begin{figure*}[ht]
\begin{minipage}{0.5 \linewidth}
 \includegraphics[width=\linewidth]{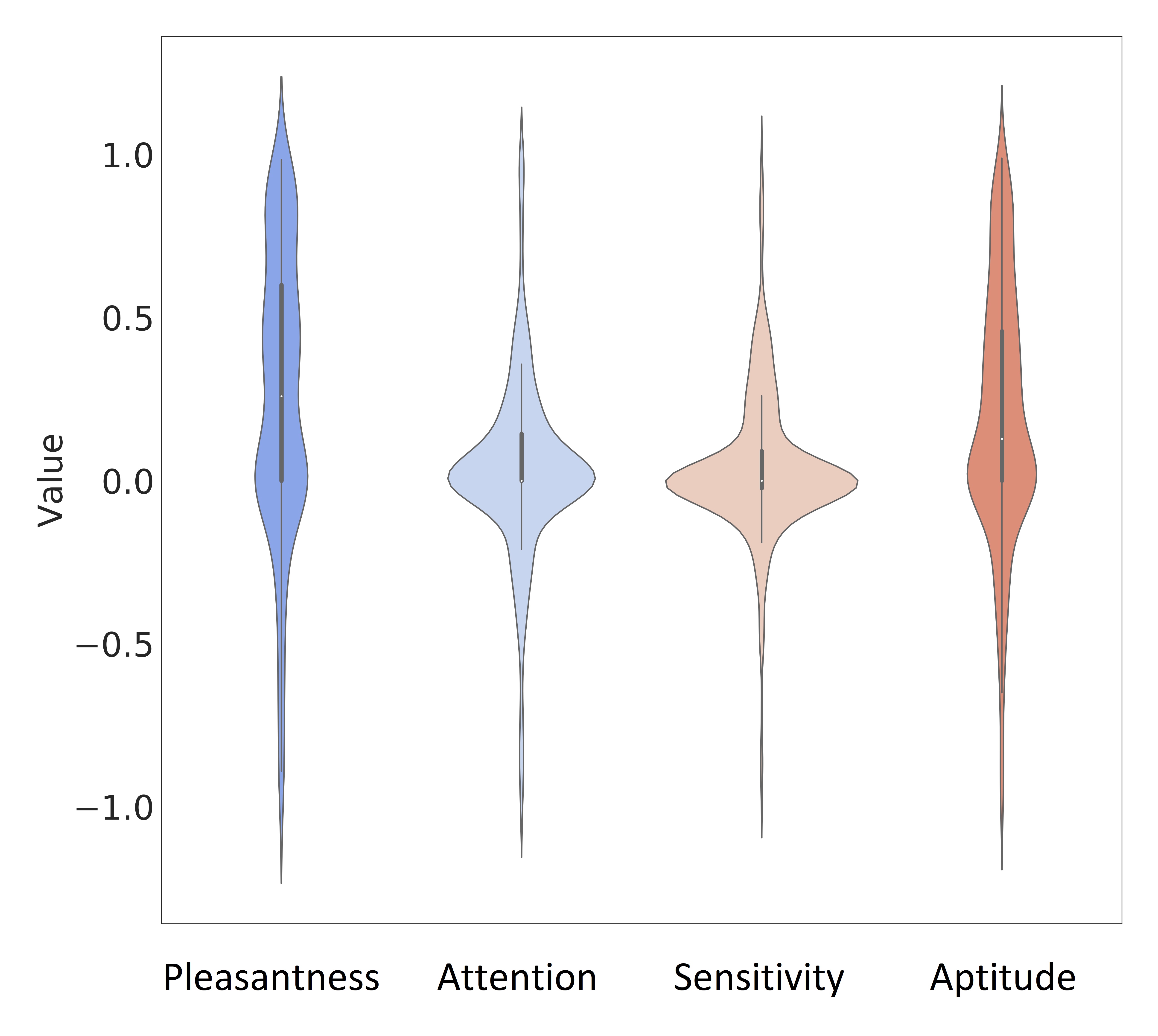}     
\end{minipage}
\hfill
\begin{minipage}{0.5 \linewidth}
 \includegraphics[width=\linewidth]{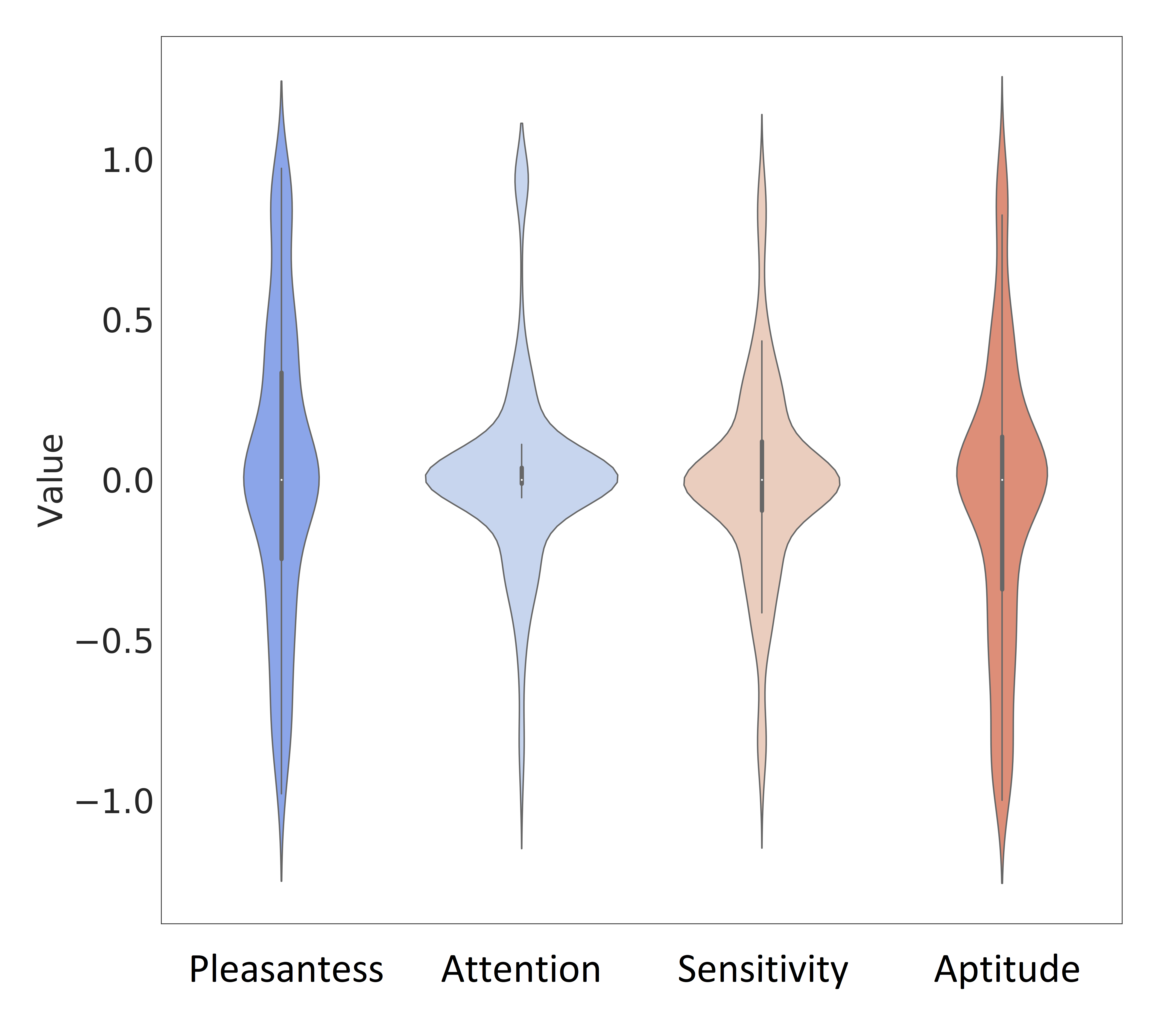}     
\end{minipage}
\caption{Distributions of the four affective dimensions from the original Hourglass of Emotions model, obtained from the SenticNet 4 framework for the dataset of English tweets. Left-hand side shows  non-hate speech tweets and right-hand side shows hate speech tweets.}
\label{fig:senticDistr}
\end{figure*}

The distributions of the affective dimensions for English tweets, separately for non-hate speech and hate speech instances, are shown in Fig. \ref{fig:senticDistr}.
While the distributions are different among the variables, the differences between the hate speech and non-hate speech distributions are not pronounced. This indicates that these  variable are not strong indicators of hate speech if used independently, but might still be useful in combination with textual features extracted by neural networks.


\subsubsection{SenticNet 6}

The revisited Hourglass of Emotions model \cite{susanto2020hourglass} is based on empirical evidence obtained in the context of sentiment analysis. Each of the four proposed baseline affective dimensions gives positive and negative perspective of one emotion: 
\begin{enumerate}
     \item  Introspection - the joy versus sadness;
    \item Temper - the calmness versus anger;
    \item Attitude - the pleasantness versus disgust, and
     \item Sensitivity - the eagerness versus fear.
\end{enumerate}

The dataset of affective dimensions was obtained using the \textit{senticnet} Python library \footnote{\url{https://pypi.org/project/senticnet/}}. We used the publicly available word level API to obtain the affective dimension values for each token separately. We averaged the affective dimension and polarity values on the level of each tweet/comment. 

\begin{figure*}[ht]
\begin{minipage}{0.5 \linewidth}
 \includegraphics[width=\linewidth]{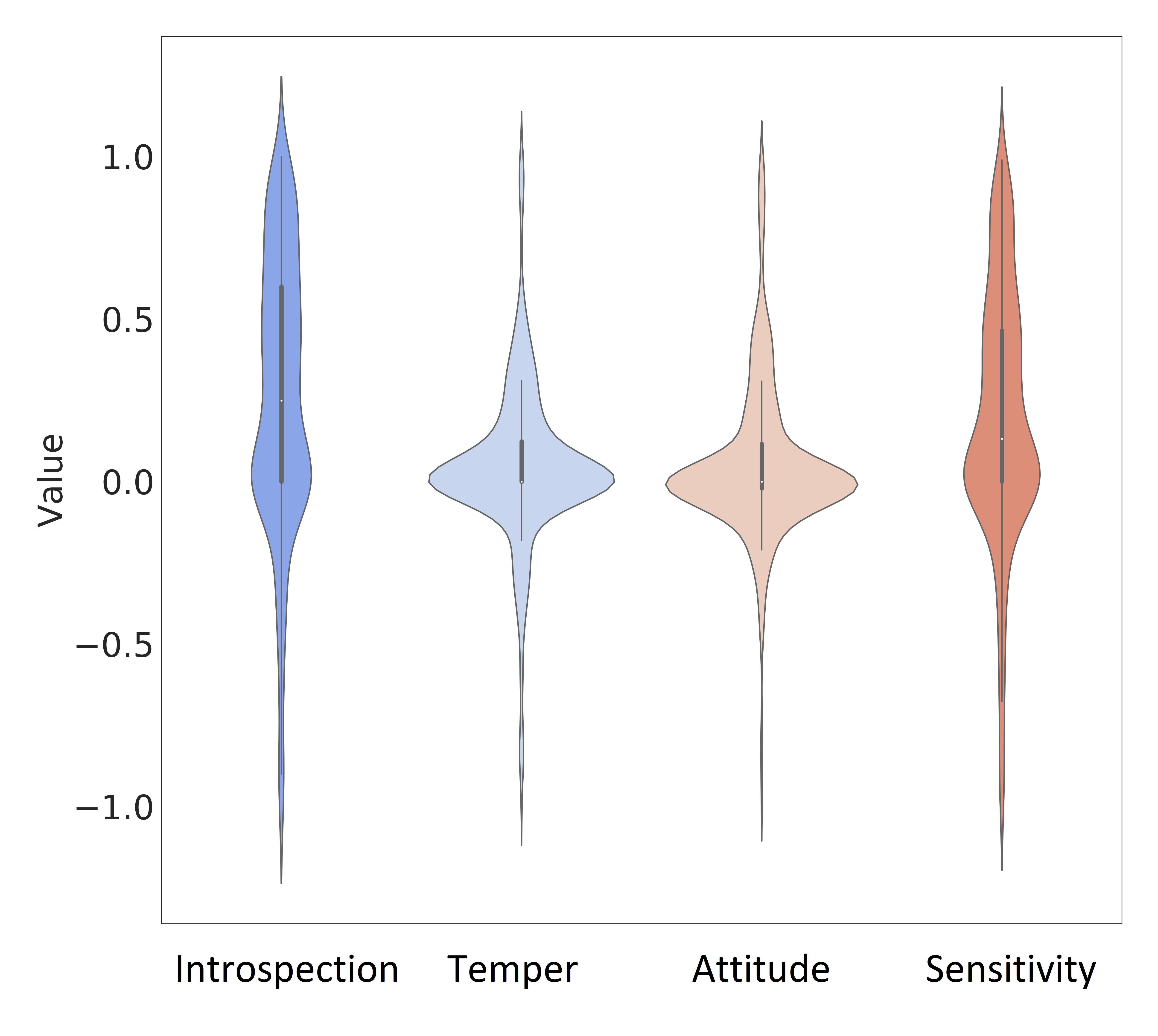}     
\end{minipage}
\hfill
\begin{minipage}{0.5 \linewidth}
 \includegraphics[width=\linewidth]{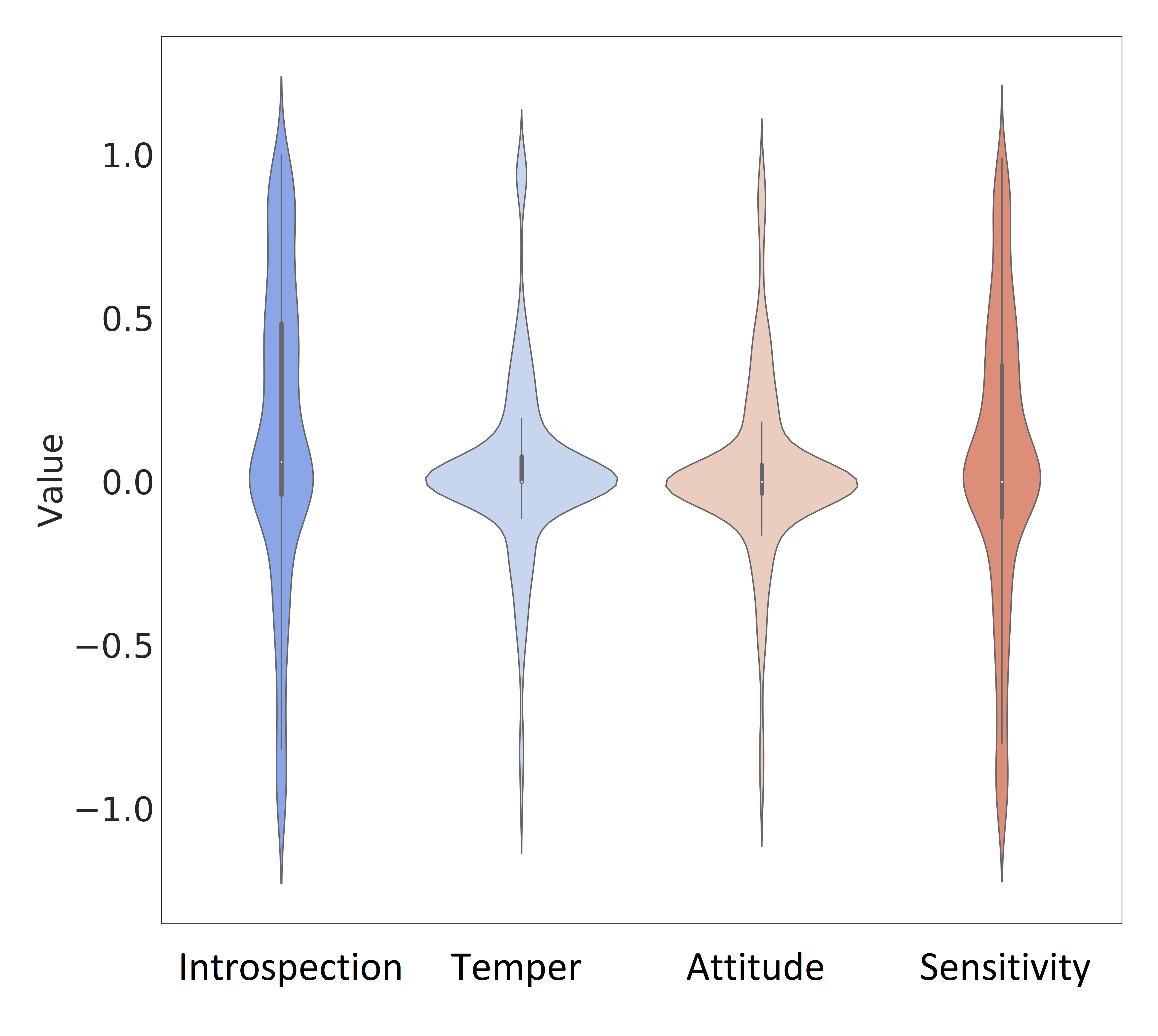}     
\end{minipage}
\caption{Distributions of the four affective dimensions from the revisited Hourglass of Emotions model, obtained from the SenticNet 6 framework for the dataset of English tweets. Left-hand side shows non-hate speech tweets and right-hand side shows hate speech tweets.}
\label{fig:senticDistr6}
\end{figure*}

We show the distributions of these new dimensions for English tweets in Fig. \ref{fig:senticDistr6}. Similarly to SenticNet 4 framework, the distributions between the hate speech and non-hate speech tweets are similar.

\subsection{Implementation of Prediction Models}
\label{sec:predictionModels}
We used three types of neural network architectures. As a baseline, we used  MCD LSTM networks \citep{miok2019prediction}, which include reliability information obtained with MCD. We compared that model with newly proposed BAN and MCD BERT. As shown in the right-most column of Table \ref{tab:models}, the input to MCD LSTM are pre-trained word embeddings: sentence encoder for English \citep{cer2018universal}, and fastText embeddings\footnote{\url{https://fasttext.cc}} for Slovene and Croatian. For the implementation of BAN, we used the Keras tokenizer\footnote{\url{https://keras.io/preprocessing/text/}}, and for MCD BERT, we used the BERT's tokenizer.

We implemented the proposed BANs\footnote{\url{https://github.com/KristianMiok/BAN}} and MCD BERT\footnote{\url{https://github.com/KristianMiok/Bayesian-BERT}} with the PyTorch library. The main hyper-parameters of the BAN architecture are the number of attention heads and the number of attention layers. The adaptive classification threshold (described in Section \ref{sec:AT}) is computed every time we evaluate the performance on the validation set. When a network makes a prediction, we deactivate all layers except the dropout layers. In this way, we maintain the variance of predictions. Each final prediction consists of a set of results obtained by several forward passes. 

Other parameters are set as follows. We use the Adamax optimizer \cite{kingma2014adam},  a variant of Adam based on infinity norm, and binary cross-entropy loss function. To automatically stop training, we use the stopping step of 10 -- if after 10 optimization steps the performance on the validation set is not improved, the training stops.

We explored the following hyperparameter tuning space: the validation percentage (size of the validation set) was varied between 5\% and 10\%. The rationale for testing different validation set sizes are relatively small datasets, therefore it is difficult to strike a good balance   between the training and validation set. Given enough data, the validation set shall be on the upper margin. 
The number of epochs was either 30 or 100, the number of hidden layers and attention heads was 1 or 2. The maximum padding of the input sequences was either 48, 32, or 64. The learning rate was either 0.001 or 0.0005, and AT was either enabled or disabled.

MCD LSTM networks consist of an embedding layer, LSTM layer, and a fully connected layer within the word2vec \cite{mikolov2013distributed} and ELMo \cite{Peters2018} embeddings. To obtain the best architectures for the LSTM and MCD LSTM models, we tested different number of units, batch sizes, dropout rates, etc. 

For BERT, we used the BERT base model in English and the multilingual BERT variant for Croatian and Slovene. We used the HuggingFace implementation\footnote{\url{https://huggingface.co/transformers/model_doc/bert.html}}.  
To combine the information from the MCD BERT and SenticNet, we generated  1000 MCD BERT predictions for each instance. We merged them with the four Sentic variables, described in Section \ref{sec:senticdata}, thus obtaining 1004 variables. This data was passed as an input to the SVM model. The process used 5-fold  cross-validation.

\subsection{Prediction Performance Evaluation Measures}
\label{sec:evaluationMeasures}
\begin{figure*}[htb]
    \centering
    \includegraphics[width = \linewidth]{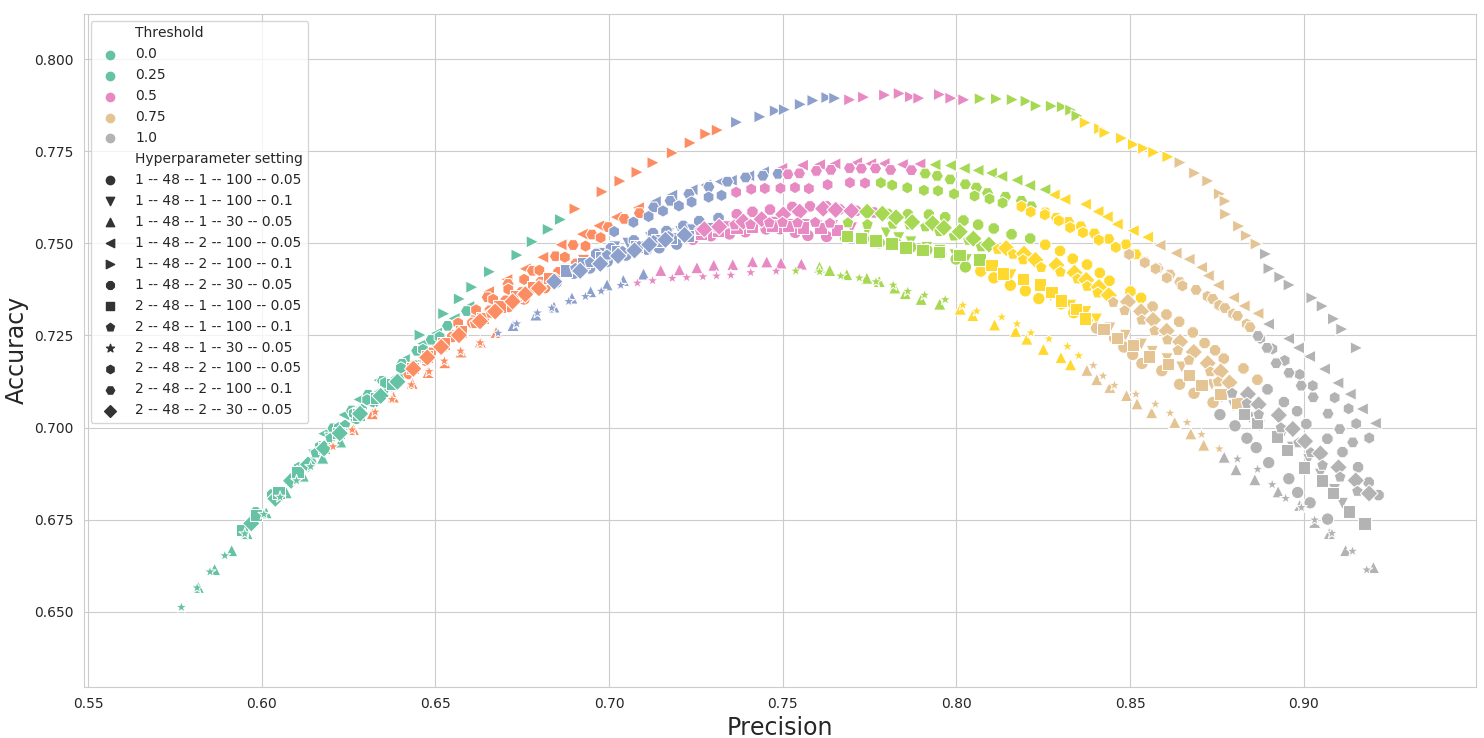}
    \caption{Trade-off between precision and accuracy across various hyper-parameters settings of BAN model. Each curve shows one set of hyper-parameters, each color depicts one decision threshold (0, 0.25, 0.5, 0.75, or 1.0). The hyper-parameters contain the number of heads, max padding, number of layers, number of epochs, and validation set ratio.}
    \label{fig:precacc}
\end{figure*}

Depending on the purpose of the prediction model, we might optimize different evaluation measures, such as classification accuracy, precision, recall, or $F_1$ score. In the hate speech detection, we want to avoid false accusations of hate speech. For that aim, we maximize precision on the validation set during training. As this could negatively affect other measure, we alter the decision threshold to achieve good precision vs. accuracy balance. In Figure \ref{fig:precacc}, we present the accuracy-precision trade-off. 

\subsection{Calibration Quality Measures}
\label{sec:calibrationEvaluation}

To measure the quality of computed calibration scores, we use the expected calibration error (ECE) \cite{guo2017calibration}.
To compute ECE, we split all $n$ predictions into $M$ equally spaced bins $B_1, B_2, \ldots, B_M$, that contain instances with prediction scores in the given bin. We sum the weighted differences between actual prediction accuracies and predicted scores over all the bins and normalizes the result with the number of instances $n$.
\begin{equation}
    ECE= \sum_{m=1}^{M}\frac{|B_m|}{n}|\textrm{accuracy}(B_m) - \textrm{score}(B_m)| 
\end{equation} 
This measure produces lower scores for better calibrated models (lower calibration error).

\section{Results}
\label{sec:results}
In this section, we present results of five sets of experiments. In Section \ref{results-calibration}, we report calibration of different prediction models, and in Section \ref{results-prediction}, their prediction performance. The comparison between the reliability of BERT and MCD BERT is presented in Section \ref{results-reliability}, while the impact of sentic features is discussed in Section \ref{results-emotions}. Finally, we present different visualizations of models' uncertainty in Section \ref{results-visualization}.

\subsection{Calibration of BAN and BERT}
\label{results-calibration}
\begin{figure}[H]
     \centering
     \includegraphics[width = 0.7 \linewidth]{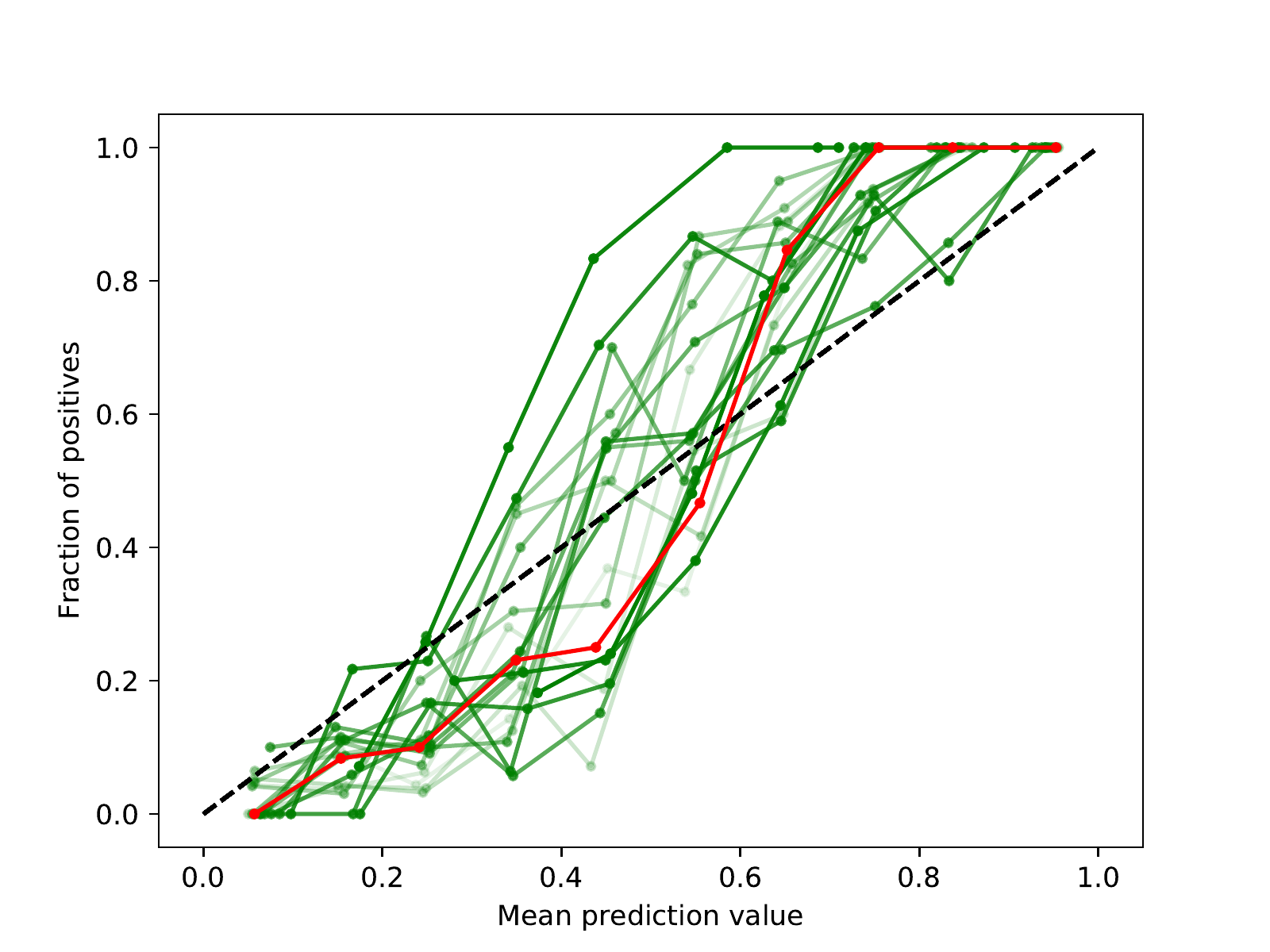}
     \caption{Calibration plot for the BAN English model after each epoch (green) based on the validation set and the best performing architecture. The transparency of the green calibrations lines decreases with the number of epochs (i.e. initial stages are the most transparent). The final calibration is in red and the dotted line shows the perfect calibration. } 
     \label{fig:calibration}
\end{figure}

Figure \ref{fig:calibration} shows how calibration of prediction scores changes during the training of BAN. The red line represents the performance of the fully trained network. It is apparent that an additional calibration is necessary -- as the perfect calibration corresponds to the dotted line. Surprisingly, some of the training iterations show better calibrated scores. This is the motivation for AT, presented in Section \ref{sec:AT}. 


In Tables \ref{calib_tab1}, \ref{calib_tab2}, and \ref{calib_tab3}, the calibration results for different calibration settings on the BAN are presented: no calibration, isotonic regression, and Platt's method. Each calibration is either combined with AT or not. For all three languages, both calibration methods improve the ECE score, and Platt's method produces the best calibration scores. The AT slightly improves the ECE score for the uncalibrated (raw) results. This is especially true for the Slovene comments where the ECE score was reduced from 0.794 to 0.621. We can conclude that the calibration using AT heuristics might not be beneficial when used in combination with the established calibration techniques (isotonic regression and Platt's method) but used exclusively.

\begin{table*}[thp]
\caption{The calibration scores of BAN with different calibration approaches on the English tweets dataset. We present average classification accuracy and F1 score with their standard deviations, computed using 5-fold cross-validation.}
\centering
\begin{tabular}{lcllr}
    \textbf{Calibration} &  \textbf{AT} &    \textbf{Accuracy} &    \textbf{F1} &  \textbf{ECE} \\
\midrule
       Raw &  False &  0.83 [0.02] &  0.82 [0.03]  &           0.547 \\
     Raw &   True &  0.83 [0.01] &  \textbf{0.83} [0.04]  &           0.539 \\
   Isotonic &  False &  \textbf{0.84} [0.01] &  0.82 [0.01] &           0.230 \\
    Isotonic &   True &  0.83 [0.01] &  0.82 [0.02] &           0.234 \\
     Platt's &  False &   \textbf{0.84} [0.02] &  0.82 [0.02] &          \textbf{0.225} \\
    Platt's &   True &  0.83 [0.01] & 0.82 [0.01] &           0.232 \\
\bottomrule
\label{calib_tab1}
\end{tabular} 
\vspace{5mm}
\caption{The calibration scores of BAN with different calibration approaches on the Croatian user news comments dataset.}
\centering
\begin{tabular}{lcllr}
    \textbf{Calibration} &  \textbf{AT} &    \textbf{Accuracy} & \textbf{F1} &  \textbf{ECE} \\
\midrule
         Raw &  False &   0.61 [0.02] &   0.47 [0.03] &           0.681 \\
         Raw &   True &  \textbf{0.62} [0.02] &  \textbf{0.50} [0.04] &           0.663 \\
    Isotonic &  False &  0.60 [0.01] &  0.49 [0.04] &           0.206 \\
    Isotonic &   True &  0.61 [0.01] &  \textbf{0.50} [0.03] &           0.206 \\
     Platt's &  False &    0.61 [0.02] &   0.48 [0.02] &           0.198 \\
    Platt's &   True &   \textbf{0.62} [0.02]  &  0.49 [0.02] &  \textbf{0.197} \\
\bottomrule
\label{calib_tab2}
\end{tabular}

\vspace{5mm}
\caption{The calibration scores of BAN with different calibration approaches on the Slovene Facebook comments  dataset.}
\begin{tabular}{lcllr}
\centering
    \textbf{Calibration} &  \textbf{AT} &    \textbf{Accuracy}  &     \textbf{F1} &  \textbf{ECE} \\
\midrule
         Raw &  False &  \textbf{0.59} [0.01]  &  0.33 [0.05] &           0.794 \\
        Raw &   True &    \textbf{0.59} [0.02]   &   0.48 [0.05] &           0.621 \\
    Isotonic &  False &  0.58 [0.02] &   0.48 [0.03] &           0.212 \\
    Isotonic &   True &  0.58 [0.02] &   \textbf{0.49} [0.03] &           0.213 \\
     Platt's &  False &   0.58 [0.03] &   0.475 [0.02] &           0.206 \\
     Platt's &   True &  \textbf{0.59} [0.02] &   0.47 [0.04] &  \textbf{0.204} \\
\bottomrule
\label{calib_tab3}
\end{tabular}
\end{table*}

\begin{figure}[htb]
    \centering
    \includegraphics[width=0.8\linewidth]{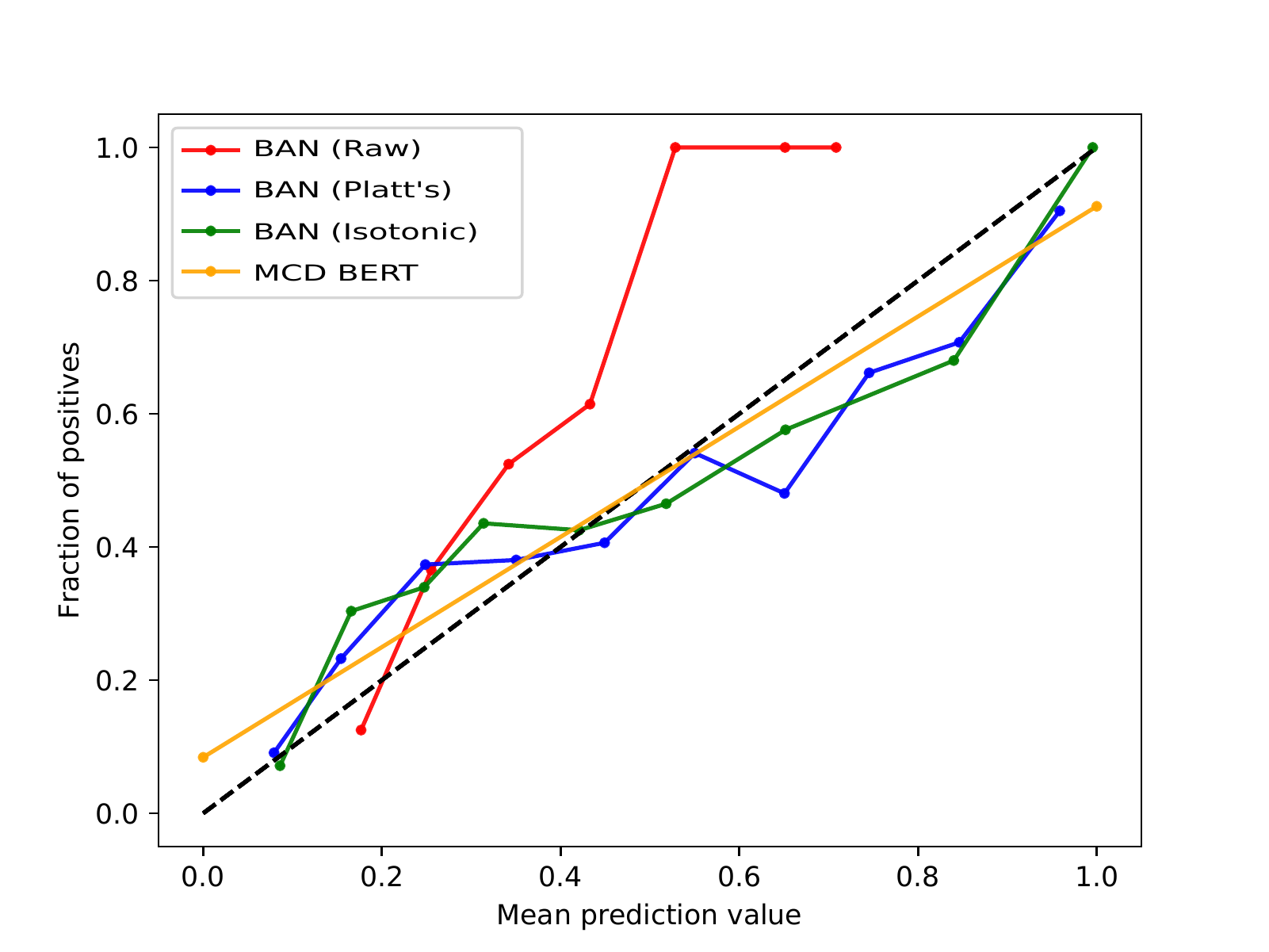}
    \caption{Calibration plots based on English test set performance for MCD BERT and BAN with different calibration algorithms.
    }
    \label{fig:CalEnglishFinal}
\end{figure}

To compare the calibration of MCD BERT with different BAN calibrations, we plotted their ECE scores in Figure \ref{fig:CalEnglishFinal}. It can be observed that calibration methods substantially improve the BAN score. However, the MCD BERT model is better calibrated even without the usage of an explicit calibration methods.       
 
\subsection{Prediction Performance}
\label{results-prediction}
We compare the predictive performance of four neural network architectures in Table \ref{tab:predictionPerformance}. MCD LSTM and BERT serve as the baselines for comparison with the proposed BAN and MCD BERT. The MCD BERT model provides the best results for all three languages. BERT models are pre-trained on large amounts of text, which makes a significant difference compared to LSTM and BAN. MCD BERT is slightly better than BERT due to its better performance for the instances where BERT is uncertain. Here, multiple predictions reduce the prediction variance.  MCD LSTM is more stable than BAN (see the standard deviation of $F_1$ score in (see the standard deviation of $F_1$ scores in Table \ref{tab:predictionPerformance}). We attribute this to the larger number of parameters in BAN and insufficient number of training instances. BERT and MCD BERT models compensate for this problem with large scale pre-training.

\begin{table*}[ht]
	\caption{Predictive performance of compared models. We present the average classification accuracy and $F_1$ score with their standard deviations (in brackets), computed using 5-fold cross-validation. The best accuracy for each language is typeset in bold. } 
	\renewcommand{\arraystretch}{1.2}
	\setlength{\tabcolsep}{4pt}
	\resizebox{\textwidth}{!}{
	\begin{tabular}{lcccccc}
	&\multicolumn{2}{c}{ \textbf{English Tweets}} & \multicolumn{2}{c}{\textbf{Croatian Comments}} & \multicolumn{2}{c}{\textbf{Slovene Comments}}\\
	 \textbf{Model} & \textbf{Accuracy}  & \textbf{F1} & \textbf{Accuracy}  & \textbf{F1} & \textbf{Accuracy}  & \textbf{F1} \\
		\hline	
\textbf{MCD LSTM} &  81.0 [1.2] &  81.9 [1.3] & 63.7 [1.0] & 51.0 [3.3] & 55.3 [0.69]  & 43.13 [0.8]    \\
\textbf{BAN} & 83.3 [1.7]  &   81.6 [3.4] & 61.4 [2.0]  & 38.1 [8.6] & 57.4 [1.7]   &  35.1 [6.3]  \\
\textbf{BERT} & 90.9 [0.7]   & 90.0 [0.7] & 70.8 [1.0] & 61.2 [1.5] & 66.4 [5.0]  &  67.8 [2.5]  \\
\textbf{MCD BERT} & \textbf{91.4 [0.7]}  & \textbf{90.4 [0.8]} & \textbf{71.5 [1.2]}  & \textbf{62.9 [1.7]}  & \textbf{68.4 [1.9]}  & \textbf{68.6 [1.6]}    \\
		\hline
	\end{tabular}}
	\label{tab:predictionPerformance}
\end{table*}

\subsection{Reliability of BERT and MCD BERT}
\label{results-reliability}
As established in Section \ref{results-calibration}, BERT models are already well-calibrated. In this section, we test if the proposed MCD BERT extension is useful beyond the advantage in predictive performance, and analyze the ability of MCD BERT to detect problematic predictions. 
For each classifier (BERT and MCD BERT), we split the tested instances into two groups, \emph{uncertain} and \emph{certain}, based on the computed prediction scores.  As BERT and MCD BERT return most of the predictions close to 0 or 1, we used the following criteria for the certainty of prediction scores. For MCD BERT, the tested instance is declared \emph{uncertain} if the variance computed on 1000 dropout predictions is greater then 0.1, otherwise it is declared \emph{certain}. As BERT returns a single prediction score, we have chosen the same number of \emph{uncertain} instances as for MCD BERT, based on the criterion that their prediction scores are farthest away from 0 or 1, i.e. they are least certain to be either hate speech or not.

\begin{table}[b]
\caption{The number and ratio of predictions where classifiers are correct/incorrect is very different for instances where BERT and MCD BERT are certain/uncertain. We use three datasets, English (ENG), Croatian (CRO), and Slovenian (SLO). }
\renewcommand{\arraystretch}{1}
\setlength{\tabcolsep}{5pt}
\label{table:chi-square}
\centering
\begin{tabular}{lcrrrr}
 &  &  \multicolumn{2}{c}{ \textbf{BERT}} & \multicolumn{2}{c}{ \textbf{MCD BERT}}\\
\textbf{Language} & \textbf{Correct}  & \textbf{Certain} & \textbf{Uncertain}  & \textbf{Certain} & \textbf{Uncertain}  \\
    \hline
\textbf{ENG}    &  \textbf{Yes}   & 880 & 31 &  891 & 24\\
     &\textbf{No} &  71 & 18 &  62 & 23 \\
   & \textbf{N/Y Ratio}   & 0.08  & 0.58 &  0.06 & 0.95 \\
 \hline
  
\textbf{CRO}     &    \textbf{Yes}&  1176 & 35  & 1053 & 152\\
      & \textbf{No} &  461 & 14 &  336 & 139 \\
      & \textbf{N/Y Ratio}   &   0.39 & 0.40 &  0.31 & 0.91 \\ \hline
       
\textbf{SLO}    & \textbf{Yes} &  576 & 28 &  537 & 55\\
     & \textbf{No} &  241 & 27& 229 & 51 \\
     & \textbf{N/Y Ratio}   & 0.42  & 0.96 &  0.42 & 0.92 \\ \hline
\end{tabular}
\end{table}

In Table \ref{table:chi-square}, we show the number of predictions where classifiers are correct/incorrect separately for instances with certain/uncertain prediction for each of the three languages. The ratio of incorrectly to correctly classified instances is significantly different between the \emph{certain} and \emph{uncertain} group, which is a strong indication that both BERT and MCD BERT correctly recognize uncertain predictions. This ratio is also much larger for MCD BERT than for BERT for the English and Croatian dataset, which testifies that the reliability of MCD BERT predictions is better. The ratio is similar for the Slovene dataset, where BERT also has a good ratio.

Using the Chi-square statistical test, we assessed the difference in correct/incorrect classifications between the \emph{certain} and \emph{uncertain} group.  For the English dataset, this difference is highly significant for both BERT and MCD BERT ($p=$1.384e-11 and 2.2e-16, respectively).  For the Croatian dataset, the p-values are 1 and 8.348e-16, meaning that we cannot rely on BERT scores to detect uncertain classifications, while the distribution returned by the MCD BERT is very informative. The p-values for the BERT and MCD BERT on Slovene are 0.0037 and 0.0002, respectively. Again, MCD BERT is much better in detecting unreliable classifications. 

The observed difference in assessment of reliability can have important practical consequences. For example, if we are faced with the re-annotation task to improve the quality of predictions, MCD BERT would choose much better borderline instances compared to BERT.


\subsection{Combining Emotional Information with MCD BERT}
\label{results-emotions}
As the experiments in Section \ref{results-prediction} show, MCD BERT is superior to other tested models on the hate speech detection task. In this section, we test if additional emotional information obtained from the SenticNet framework can complement the information about the hate speech extracted by the MCD BERT model and further improve its performance. We merge the affective dimensions computed based on SenticNet 4 and SenticNet 6 with the output vector of MCD BERT predictions, described in Section \ref{sec:MCDBERT}. Additionally, we investigate if the emotional information can help in the interpretation of trained hate speech detectors. 

\begin{figure}[bth]
    \centering
    \includegraphics[width=1\linewidth]{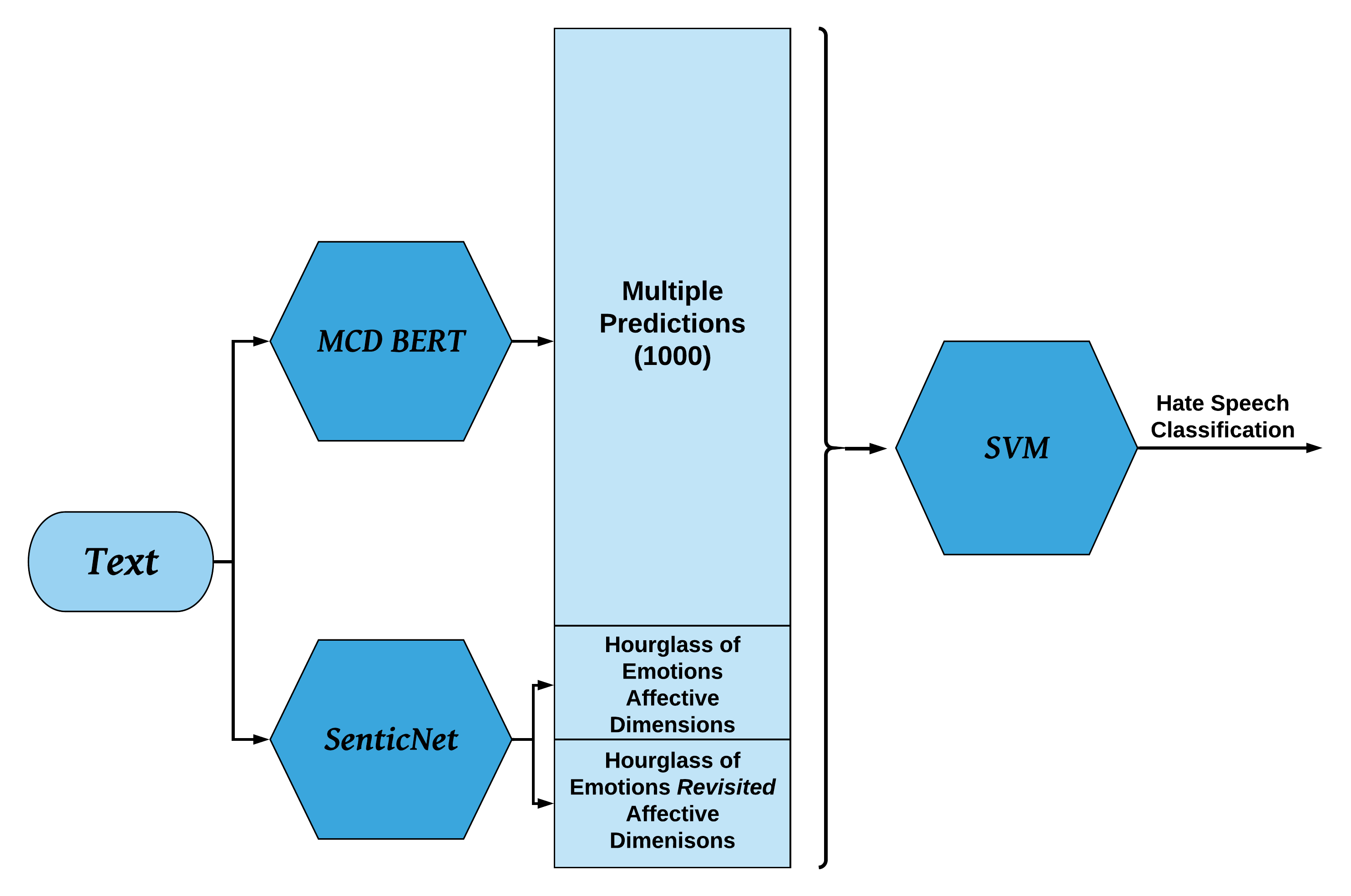}
    \caption{A diagram of merging MCD BERT predictions with the emotional information based on SenticNet 4 and SenticNet 6 frameworks. The concatenated vector is an input to the final SVM classifier that predicts the hate speech.}
    \label{fig:comb}
\end{figure}

\begin{figure}[htb]
    \centering
    \includegraphics[width=0.7\linewidth]{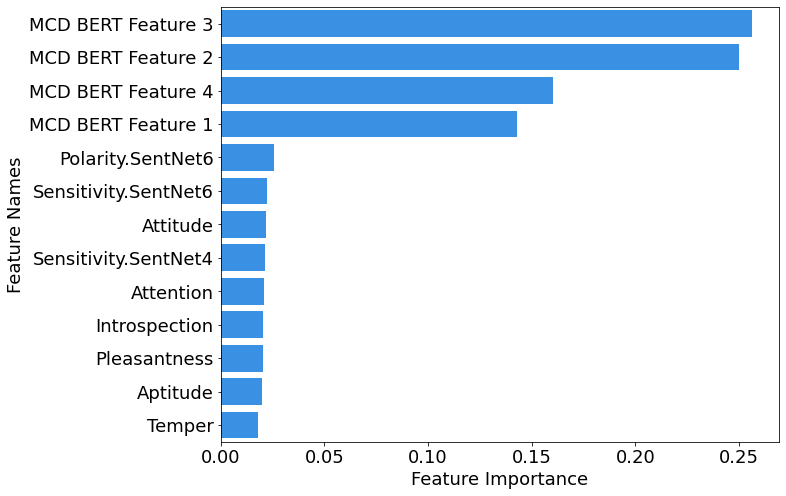}
    \caption{Feature importance scores according to the random forest algorithm. We show scores of 8 affective dimensions extracted from the SenticNet 4 and SenticNet 6 frameworks, as well as five most important attributes generated by the MCD BERT model.
    }
    \label{fig:Sentic0}
\end{figure}

For the evaluation, we use 5-fold cross-validation. In each iteration, we combine the predictions from MCD BERT (1000 of them, sorted in ascending order) with the affective dimensions from the original and revised variant of the Hourglass of Emotions models as depicted in Figure \ref{fig:comb}. We obtain four affective dimensions from the original Hourglass of Emotions model (pleasantness, aptitude, sensitivity, and attention), and four from the revisited model (introspection, sensitivity, temper, and attitude). 
Using the dataset obtained in this way, we train the SVM model to predict the hate speech. According to the results in Table \ref{tab:combination}, the additional information does not improve the hate speech detection. The same conclusions can be drawn from Figure \ref{fig:Sentic0}, where we plot the scores assigned to the used features by the random forest algorithm \cite{Breiman01}. This learning algorithm can detect feature dependencies that affect the prediction variable. Thus, the results show that SVM and random forest cannot detect any pronounced interactions between affective dimensions and MCD BERT predictions that would impact the hate speech classification.

The results show that introducing knowledge regarding emotional content
after the predictions are done can not improve the performance. However,
according to the authors of the Hourglass of Emotions revisited model \citep{susanto2020hourglass}, the full sentence model introduced in SenticNet 6 \citep{cambria2020senticnet} can provide superior text classification results on problems involving emotions. Thus, the layers that can capture emotional information from the text should be build within the prediction model architecture. Introducing uncertainty component in such architecture remains an interesting direction for further research. 

\begin{table*}[htb]
	\caption{Predictive performance of the MCD BERT model and the SVM model trained on the output features of MCD BERT and affective dimensions from the two Hourglass of Emotions models for the English tweets dataset. 
	} 
	\centering
	\renewcommand{\arraystretch}{1.2}
	\setlength{\tabcolsep}{12pt}
	\begin{tabular}{lcc}
	 \textbf{Model} & \textbf{Accuracy}  & \textbf{F1}\\
		\hline			
\textbf{MCD BERT} & \textbf{91.4 [0.7]}  & \textbf{90.4 [0.8]}  \\
\textbf{MCD BERT + \textit{SenticNet 4} + \textit{SenticNet 6}} & \textbf{91.4 [0.5]}  & \textbf{90.5 [0.9]}   \\
		\hline
	\end{tabular}
	\label{tab:combination}
\end{table*}

To better understand the emotions involved in hate speech problem, we further investigated the relation  between the affective dimensions of the two Hourglass of Emotions models (original and revisited) and the hate speech prediction probabilities of MCD BERT, separately for the non-hate speech and hate speech English tweets. 

\begin{figure}[htb]
    \centering
    \includegraphics[width=10cm, height=3cm]{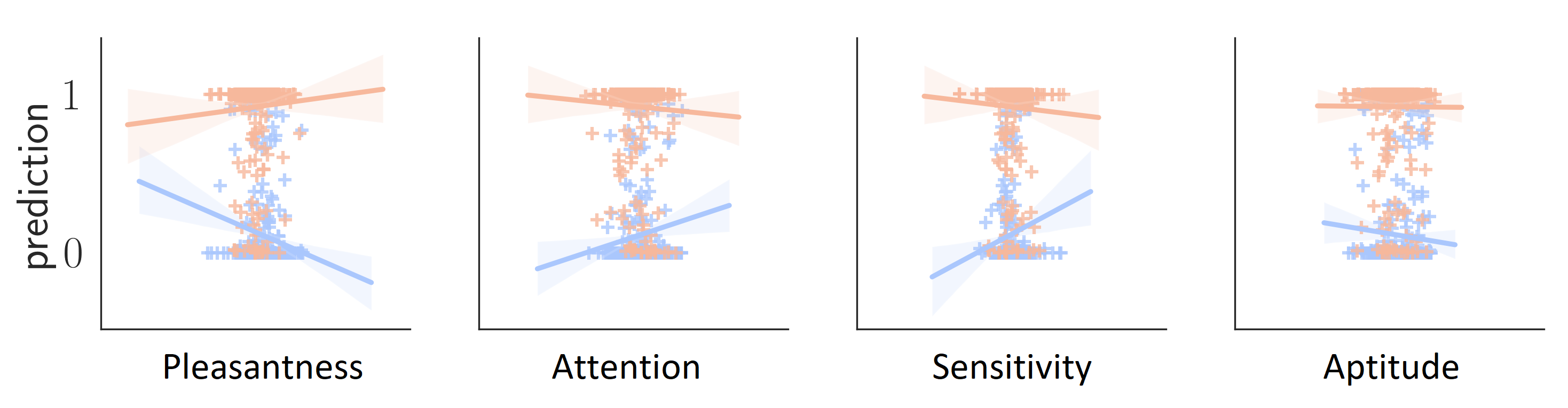}
        \includegraphics[width=10cm, height=3cm]{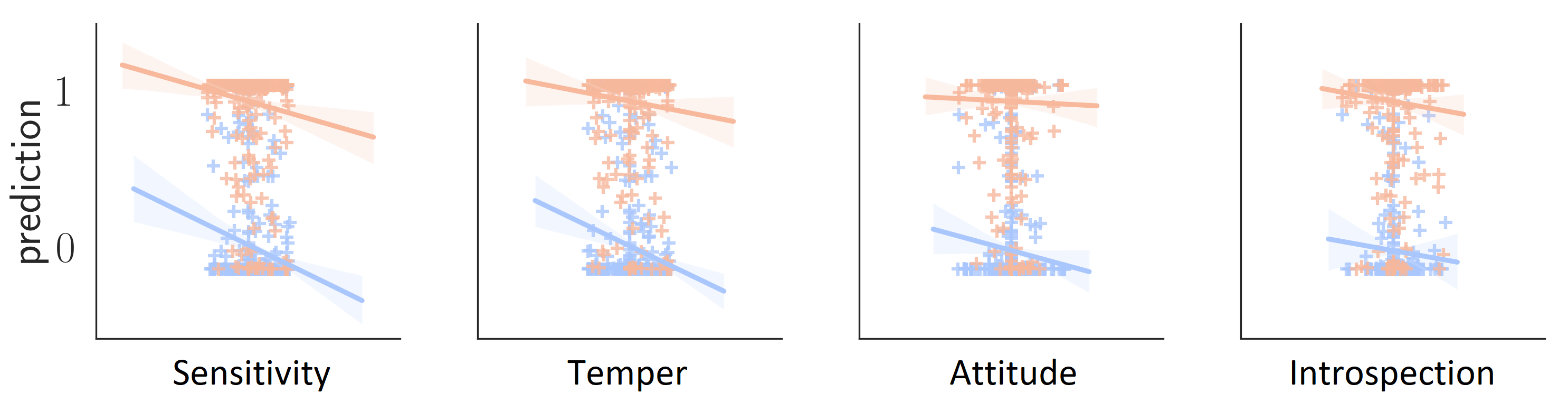}
    \caption{Relationship between prediction probability of MCD BERT and the  Hourglass of Emotions affective dimensions. Original affective dimensions are shown in the top line, while revisited dimensions are shown in the bottom line of graphs.}
    \label{fig:Sentic2}
\end{figure}

The top line of Figure \ref{fig:Sentic2} shows results for the affective dimensions of the original Hourglass of Emotions models (pleasantness, attention, sensitivity, and aptitude). The top parts of graphs  show that linear regression lines (in orange) for the hate speech are almost horizontal, so there is no significant correlation between the predicted probability of hate speech obtained with MCD BERT and affective dimensions. In contrast, the correlation between the predicted probability and affective dimensions for the non-hate speech tweets is significant, as the blue regression lines at the bottom parts of the graphs in the top row show. Both \textit{attention} and \textit{sensitivity} have positive correlation with the hate speech prediction probability. This is in accordance with the conclusions of the original Hourglass of Emotions model that high \textit{attention} and \textit{sensitivity} lead to aggressiveness (Figure 5 in \cite{cambria2012org}).

The bottom line of Figure \ref{fig:Sentic2} shows the affective dimensions of the revisited Hourglass of Emotions model (sensitivity, temper, attitude and introspection). These affective dimensions are all negatively correlated with the non-hate speech. It can be observed that there is also a slight negative correlation between the affective dimensions and hate speech probabilities, especially for sensitivity and temper. Thus, tweets that contain dominantly positive emotions have a low probability of being hate speech which is in accordance with the results presented by \citet{susanto2020hourglass}.

\subsection{Visualization of Uncertainty}
\label{results-visualization}

\begin{figure*}[ht]
\begin{minipage}{0.2 \linewidth}
 \includegraphics[width=\linewidth]{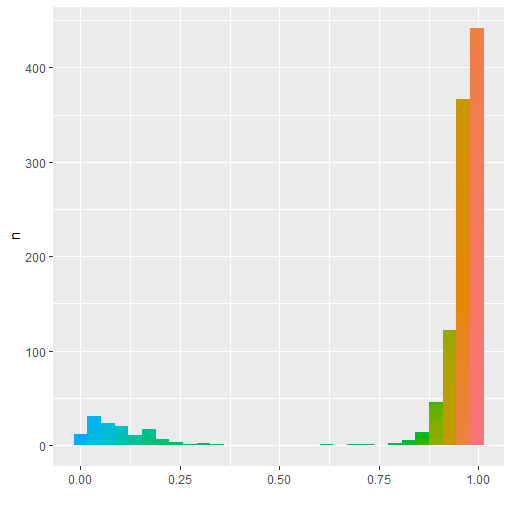}     
\end{minipage}
\hfill
\begin{minipage}{0.2 \linewidth}
 \includegraphics[width=\linewidth]{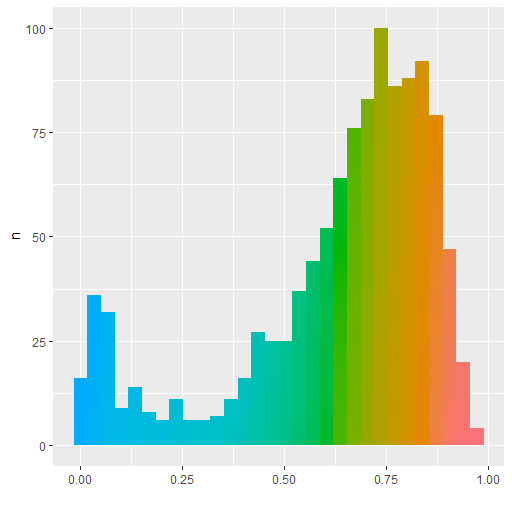}     
\end{minipage}
\hfill
\begin{minipage}{0.2 \linewidth}
 \includegraphics[width=\linewidth]{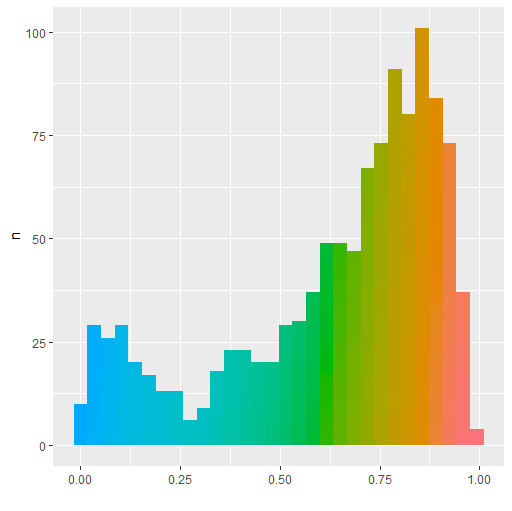}     
\end{minipage}
\hfill
\begin{minipage}{0.2 \linewidth}
 \includegraphics[width=\linewidth]{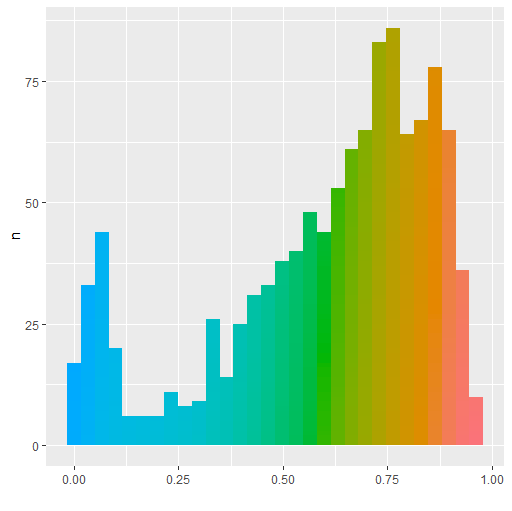}     
\end{minipage}

\begin{minipage}{0.2 \linewidth}
 \includegraphics[width=\linewidth]{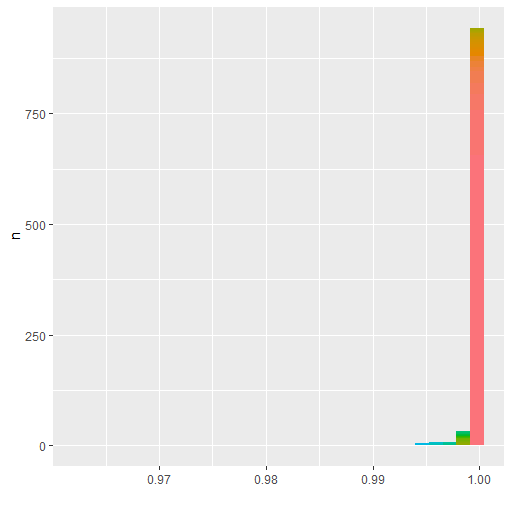}     
\end{minipage}
\hfill
\begin{minipage}{0.2 \linewidth}
 \includegraphics[width=\linewidth]{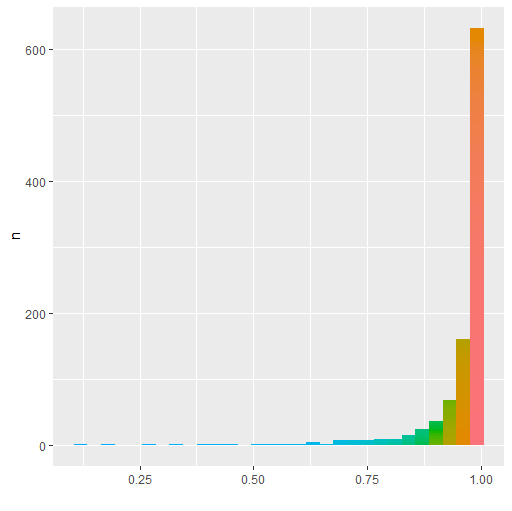}     
\end{minipage}
\hfill
\begin{minipage}{0.2 \linewidth}
 \includegraphics[width=\linewidth]{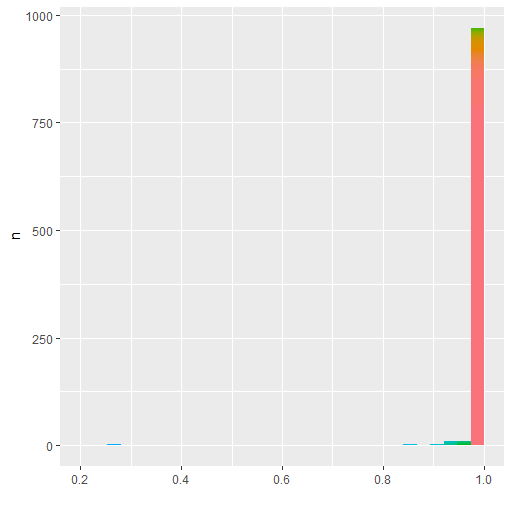}     
\end{minipage}
\hfill
\begin{minipage}{0.2 \linewidth}
 \includegraphics[width=\linewidth]{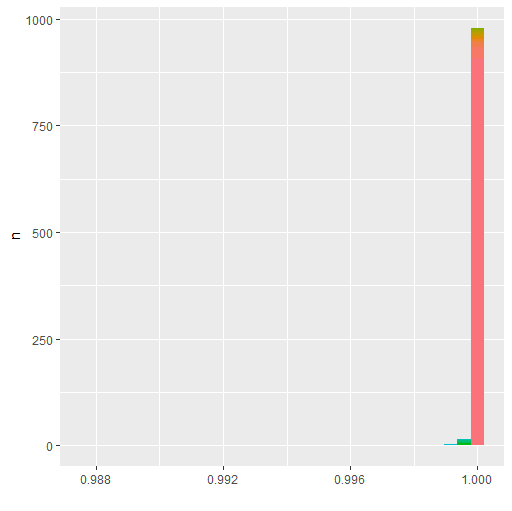}     
\end{minipage}

\begin{minipage}{0.2 \linewidth}
 \includegraphics[width=\linewidth]{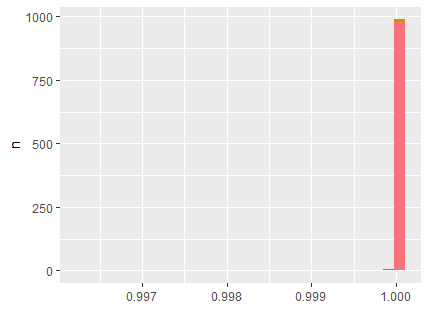}     
\end{minipage}
\hfill
\begin{minipage}{0.2 \linewidth}
 \includegraphics[width=\linewidth]{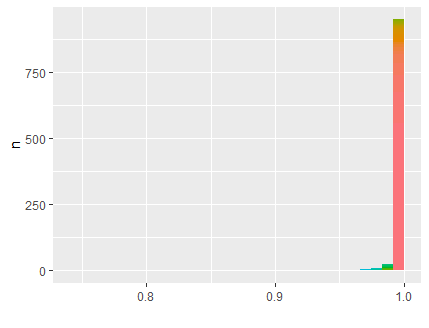}     
\end{minipage}
\hfill
\begin{minipage}{0.2 \linewidth}
 \includegraphics[width=\linewidth]{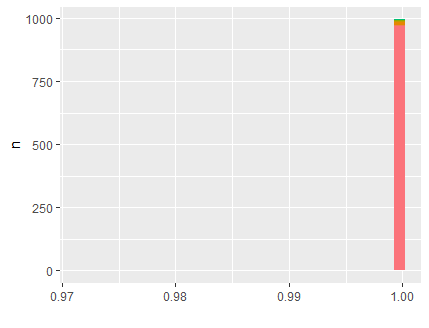}     
\end{minipage}
\hfill
\begin{minipage}{0.2 \linewidth}
 \includegraphics[width=\linewidth]{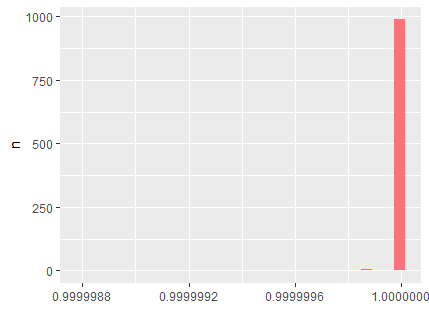}     
\end{minipage}

\begin{minipage}{0.2 \linewidth}
 \includegraphics[width=\linewidth]{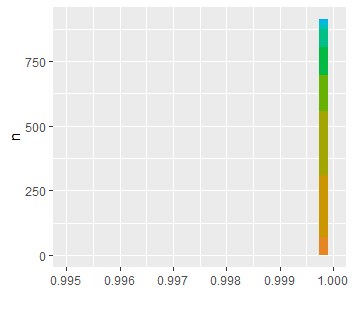}     
\textbf{tweet1:} @user you might be a libtard if... \#libtard  \#sjw \#liberal \#politics  
\end{minipage} 
\hfill
\begin{minipage}{0.22 \linewidth}
 \includegraphics[width=\linewidth]{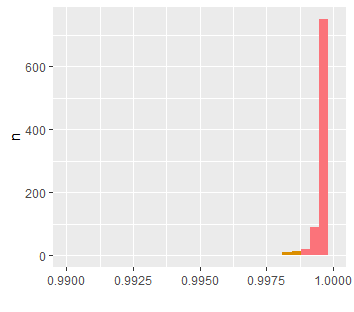}     
\textbf{tweet2:} \#paid \#kkk to \#fabricate \#stories to push the producers \#narrative  \#cancel 
\end{minipage}
\hfill
\begin{minipage}{0.22 \linewidth}
 \includegraphics[width=\linewidth]{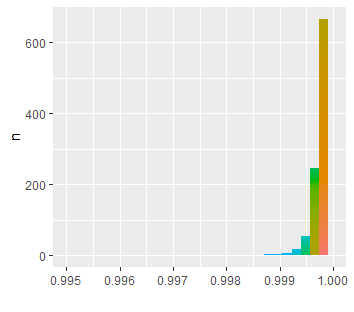}     
\textbf{tweet3:} carl paladino, trump ally, wishes obama dead of mad cow disease  
\end{minipage}
\hfill
\begin{minipage}{0.22 \linewidth}
 \includegraphics[width=\linewidth]{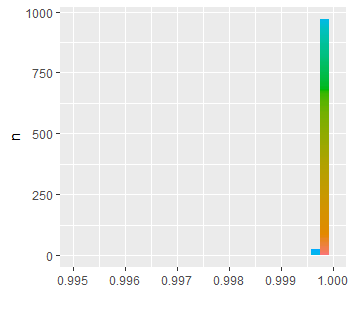}     
\textbf{tweet4:} thought factory: left-right polarisation! \#trump \#uselections2016 \#leadership \#politics  \#brexit \#blm   
\end{minipage}
\caption{Distributions of prediction scores for a few \emph{correctly} classified English instances. We show histograms for MCD LSTM (first row), BAN with $30\%$ dropout (second row), BAN with $10\%$ dropout (third row), and MCD BERT (fourth row). Each tweet is shown in a separate column.}
\label{fig:histogramsCorrect}
\end{figure*}

\begin{figure*}[ht!]
\begin{minipage}{0.2 \linewidth}
 \includegraphics[width=\linewidth]{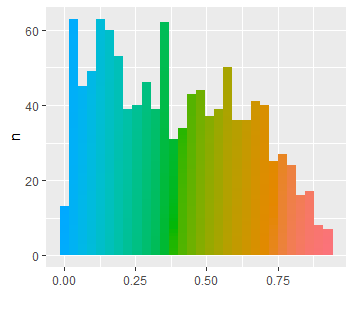}     
\end{minipage}
\hfill
\begin{minipage}{0.2 \linewidth}
 \includegraphics[width=\linewidth]{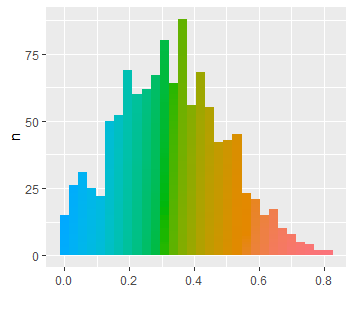}     
\end{minipage}
\hfill
\begin{minipage}{0.2 \linewidth}
 \includegraphics[width=\linewidth]{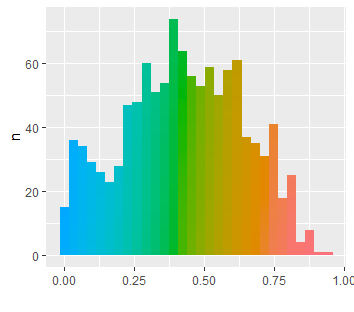}     
\end{minipage}
\hfill
\begin{minipage}{0.2 \linewidth}
 \includegraphics[width=\linewidth]{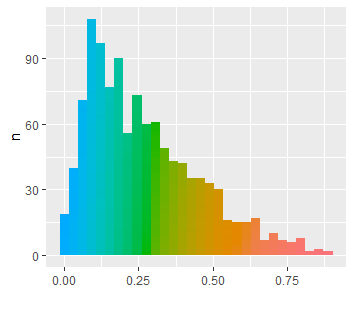}     
\end{minipage}

\begin{minipage}{0.2 \linewidth}
 \includegraphics[width=\linewidth]{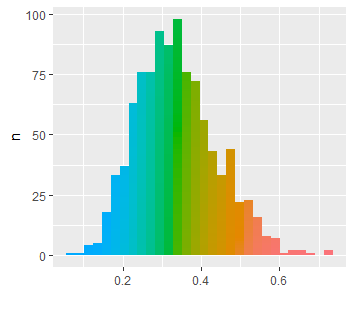}     
\end{minipage}
\hfill
\begin{minipage}{0.2 \linewidth}
 \includegraphics[width=\linewidth]{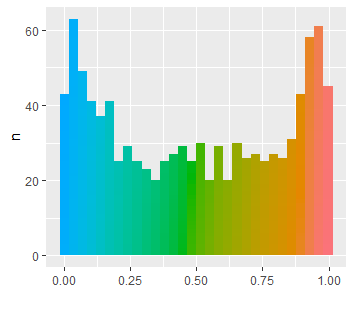}     
\end{minipage}
\hfill
\begin{minipage}{0.2 \linewidth}
 \includegraphics[width=\linewidth]{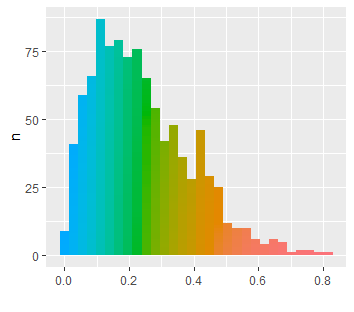}     
\end{minipage}
\hfill
\begin{minipage}{0.2 \linewidth}
 \includegraphics[width=\linewidth]{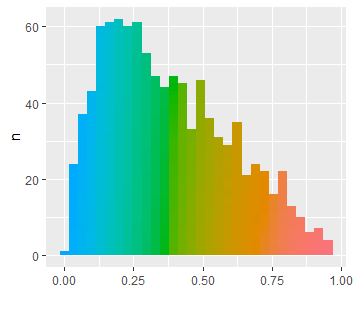}     
\end{minipage}
\begin{minipage}{0.2 \linewidth}
 \includegraphics[width=\linewidth]{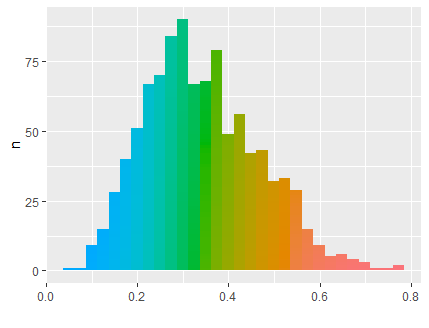}     
\end{minipage}
\hfill
\begin{minipage}{0.2 \linewidth}
 \includegraphics[width=\linewidth]{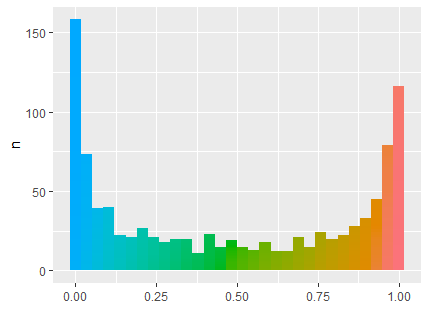}     
\end{minipage}
\hfill
\begin{minipage}{0.2 \linewidth}
 \includegraphics[width=\linewidth]{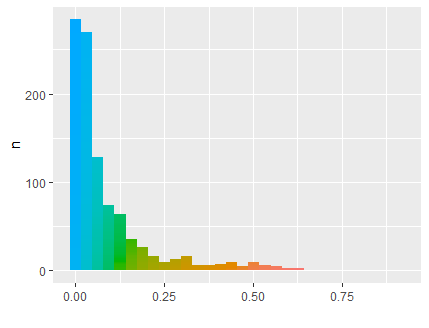}     
\end{minipage}
\hfill
\begin{minipage}{0.2 \linewidth}
 \includegraphics[width=\linewidth]{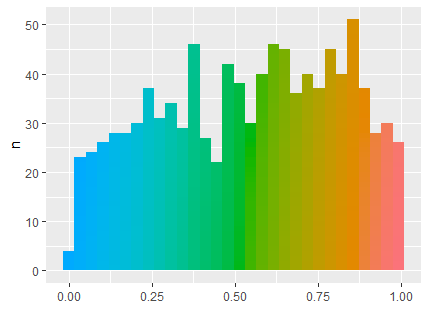}     
\end{minipage}
\begin{minipage}{0.2 \linewidth}
 \includegraphics[width=\linewidth]{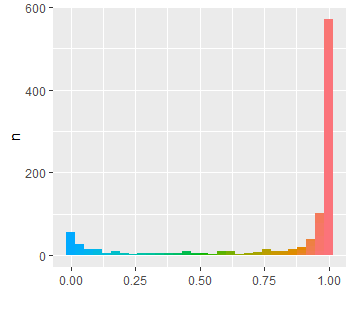}     
\textbf{tweet5:} what over-rode journalists' integrity was greed and ambition, along with a total absence of courage (true label = \textcolor{red}{hate speech})
\end{minipage} 
\hfill
\begin{minipage}{0.22 \linewidth}
 \includegraphics[width=\linewidth]{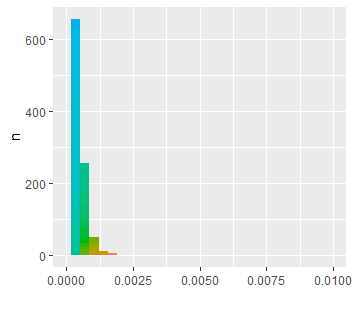}     
\textbf{tweet6:} there is a future nina and right now you are pretty much throwing yours away. (true label = \textcolor{blue}{non-hate speech})
\end{minipage}
\hfill
\begin{minipage}{0.22 \linewidth}
 \includegraphics[width=\linewidth]{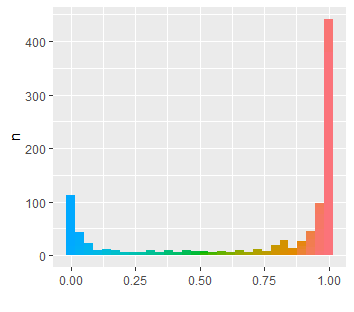}     
\textbf{tweet7:} @user @user someone really respects eu guidelines!!! (true label = \textcolor{blue}{non-hate speech})
\end{minipage}
\hfill
\begin{minipage}{0.22 \linewidth}
 \includegraphics[width=\linewidth]{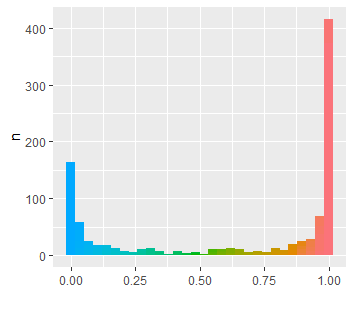}     
\textbf{tweet8:} the scars left by us waime camps  mt @user (true label = \textcolor{red}{hate speech})
\end{minipage}
\caption{Distributions of prediction scores for a few \emph{incorrectly} classified English instances. We show histograms for MCD LSTM (first row), BAN with $30\%$ dropout (second row), BAN with $10\%$ dropout (third row), and MCD BERT (fourth row). Each tweet is shown in a separate column. }
\label{fig:histogramsWrong}
\end{figure*}

Obtaining multiple predictions for a specific instance can improve understanding of the final prediction. We used the mean of the distribution to estimate the probability. The variance informs us about the spread and certainty of predictions. We can inspect the actual distribution of prediction scores with histogram plots, as illustrated in Figure \ref{fig:histogramsCorrect} for a few correctly classified instances from the English dataset, and on Figure \ref{fig:histogramsWrong} for a few misclassified instances. We analyze distributions produced by the MCD LSTM baseline, BAN with 10\% and 30\% dropout, and MCD BERT. 

Histograms in Figures \ref{fig:histogramsCorrect} and \ref{fig:histogramsWrong}  visually display the prediction certainty for the specific instances. We notice that MCD BERT's predictions are always close to 0 or 1, especially when the model seems certain of the prediction. BAN with 10\% dropout provides a similar spread of values as MCD BERT. This is expected as BERT is also pre-trained with 10\% dropout. However, 30 \% of dropout in BAN results in a much larger spread of predictions for instances where BAN is uncertain. Note that the results of MCD BERT are concentrated in a much narrower interval compared to MCD LSTM and BAN.

\begin{figure*}[ht!]
    \centering
    \includegraphics[width = 0.75\linewidth]{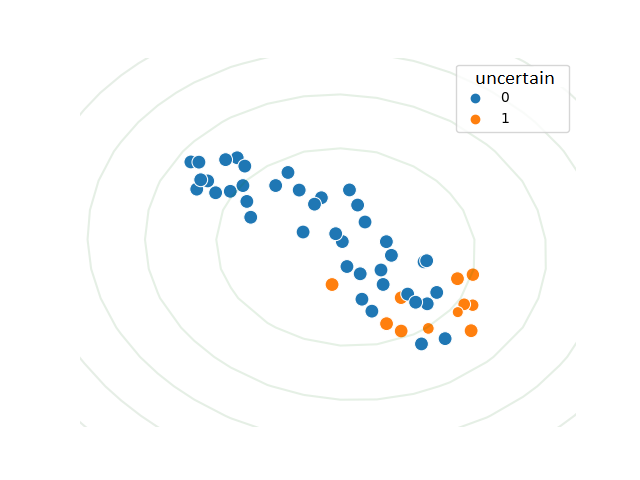}
    \caption{Visualization of  100 test tweets projected into two dimensional space by the UMP method. Tweets whose classifications seem certain are colored in blue while tweets with uncertain classification are shown in orange. We can observe clustering of uncertain tweets. }
    \label{fig:probviz1}
    \includegraphics[width = 0.75\linewidth]{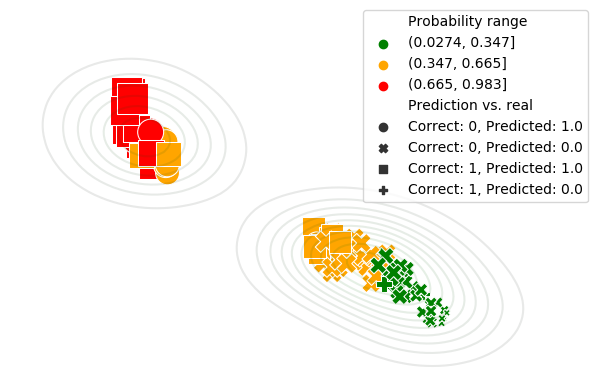}
    \caption{Visualization of the probability space for 100 tweets from the test set. The instances are colored green, yellow, or red, depending on the mean probability of the 1000 predictions. Predictions with high confidence form an isolated part in the probability space.}
    \label{fig:probviz2}
\end{figure*}

While visualizations of prediction distributions for individual instance (see Figures \ref{fig:histogramsCorrect} and \ref{fig:histogramsWrong}) are useful in the assessment of their prediction reliability, we also aggregate results over multiple instances to understand more general reliability phenomena. Following \cite{miok2019prediction}, we visualize the embeddings of the prediction distributions. The idea of this visualization is to detect and identify clusters of certain and uncertain classifications. First, we obtain many predictions (1,000 in our experiments)  for each instance. The space of prediction distributions across instances is embedded into two dimensions by the Uniform Manifold Projections method (UMP) \cite{mcinnes2018umap}. In this way, we obtain a two-dimensional space corresponding to the initial 1,000 dimensional space of prediction distributions. Next, we use the Gaussian kernel estimation to identify equivalent regions and connect them with closed curves. Finally, the shapes and sizes of individual predictions are chosen based on their classification error and certainty of predictions. The goal of this visualization is to discover structures within the space of probability distributions, possibly offering insights into the drawbacks and limitations of the analyzed classifier. The resulting visualizations are shown in Figures~\ref{fig:probviz1} and ~\ref{fig:probviz2}. In Figure~\ref{fig:probviz1}, the plot displays the position of certain and uncertain test set instances in the embedded space of distributions, while in Figure~\ref{fig:probviz2} the differences are based on the mean of predicted probability scores.

In both Figures, \ref{fig:probviz1} and \ref{fig:probviz2}, the probability space is distinctly separated into two components, indicating that there are predictions for which the neural network is certain (and were correctly classified). However, for some predictions, especially  non hate speech instances, the model is less certain (albeit still correct). The two visualizations demonstrate how the probability space is split into distinct components for a trained neural network. The visualizations also shows problematic predictions, allowing their identification and potentially facilitating the debugging process for developers (e.g., an inspection of convergence).

\section{Conclusions and Future Work}
\label{sec:conclusions}
In real world scenarios, an automatic detection of hate speech requires high precision and reliable decisions. Wrong classifications can lower the level of democratic debate and damage freedom of speech. In technological terms, NLP is witnessing a switch from RNNs with pre-trained word embeddings (such as LSTM with fastText) to large pre-trained transformer models (such as BERT).

We proposed to use the MCD in the attention layers of transformer neural networks, and to unfreeze dropout layers also during the prediction phase. This resulted in two new architectures, BAN and MCD BERT. The BAN models are transformer networks trained from scratch, using dropout in both training and prediction phase. MCD BERT uses pretrained BERT model and uses dropout during fine-tuning and prediction phase. We have shown that these approaches are useful for estimation of prediction uncertainty. MCD BERT significantly improves the prediction performance in the hate speech detection task. Its pre-training extracts useful information about the language use that can be successfully exploited in the fine-tuning to a specific problem. BANs, trained from scratch, are not competitive with this. We also empirically investigated the calibration of BAN and MCD BERT. The results show that MCD BERT is much better calibrated than BAN. 

Multiple predictions obtained from MCD BERT not only produce better predictive performance compared to BERT, but also provide better reliability information. The visualizations based on them enable detection of less certain decisions and can help moderators or annotators to focus on uncertain instances.

In line with the recent research showing that the affective information available in the SenticNet 6 framework provides favorable  results in the sentiment analysis \citep{susanto2020hourglass}, we tested this information on the hate speech detection task. We combined affective dimensions from the original and revisited Hourglass of Emotions models with predictions generated by the MCD BERT model. While our results do not show any improvement in predictive performance, we believe inclusion of affective information should be incorporated within the prediction model together with possibility of obtaining prediction uncertainty. Thus, we see an opportunity for further work in this area by introducing BERT-based uncertainty estimated into full sentence models from the SenticNet 6 framework. Nevertheless, the predictions of the MCD BERT model confirm the findings of the Hourglass of Emotions model. The affective dimensions of the Hourglass of Emotions model are correlated with the non-hate speech probabilities returned by the MCD BERT, and can potentially explain emotions involved in the hate speech. Breaking down a complex offensive language to fundamental emotions can bring interesting insights into the hate speech problem.  

In future work, we aim to adapt other Bayesian approaches, such as SWAG, to transformer networks. Reliability enhanced classifications could be used in many other domains, such as machine translation. One of the tasks where Bayesian text classification can be particularly useful is semi-supervised learning, which iteratively expands an initial small set of manually labeled instances with the most reliably classified instances. Data re-annotation is another example where reliability scores can be of great use. An initial pilot study on Croatian comment filtering showed that human annotators decide mostly based on the observed keywords and lack the time to detect more subtle expressions of offensive content. These circumstances result in low quality of the resulting datasets and demand their reannotation. Using the reliability scores of the proposed MCD BERT,  one could significantly reduce the amount of reannotation and focus on genuinely difficult and borderline cases where prediction models may err. 
\\
\\
\textbf{Funding Information} Marko Robnik-\v Sikonja received the financial support from the Slovenian Research Agency through core research programme P6-0411. All the authors except Daniela Zaharie have received funding from the European Union’s Horizon 2020 research and innovation programme under grant agreement No 825153 (EMBEDDIA, Cross-Lingual Embeddings for Less-Represented Languages in European News Media).
\\
\\
\textbf{Compliance with Ethical Standards}
\\
\\
\textbf{Conflict of Interest} The authors declare that they have no conflict of interest. 
\\ 
\\
\textbf{Ethical Approval} This article does not contain any studies with human participants or animals performed by any of the authors.
\\
\\
\textbf{Informed Consent}
Informed consent was not required as no humans
or animals were involved.


\bibliographystyle{spbasic}
\bibliography{learn}

\end{document}